\documentclass[preprints,article,accept,moreauthors,pdftex,10pt]{mdpi}
\firstpage{1}
\pubvolume{xx}
\issuenum{1}
\articlenumber{5}
\pubyear{2018}
\copyrightyear{2018}
\history{Received: date; Accepted: date; Published: date}
\usepackage{caption}
\usepackage{subcaption}
\usepackage{graphicx}
\usepackage{CJKutf8}
\usepackage{CJK}

\Title{Highly Dilute Gas Flows Through A  Non-Isothermal Planar Micro-Channel  }

\Author{Shiying Cai{\begin{CJK*}{UTF8}{gbsn}
(蔡世英)\end{CJK*}$^1$}, 
Chunpei Cai{\begin{CJK*}{UTF8}{gbsn}
(蔡春培)\end{CJK*}$^1$} and Jun Li{\begin{CJK*}{UTF8}{gbsn}
（李军)\end{CJK*}$^2$} }
\AuthorNames{Shiying Cai, Chunpei Cai and Jun Li}
\address{%
$^{1}$ \quad Department of Mechanical Engineering-Engineering Mechanics, Michigan Technological University, Houghton,  1400 Townsend Dr., Michigan 49931, USA \\
$^{2}$ \quad Center for Integrative Petroleum Research, College of Petroleum Engineering \& Geosciences,  King Fahd University of Petroleum and Minerals, Dhahran 31261, Saudi Arabia}

\corres{Emails: ccai@mtu.edu; junli@kfupm.edu.sa}

\abstract{This paper reports theoretical and numerical investigations on free molecular gas flows through micro-channels. Both diffusely and specularly reflective channel surfaces are considered. Gaskinetic methods are adopted to develop the analytical solutions for surface and flowfield properties.  The crucial steps include  constructing the velocity distribution functions (VDFs) for points at the plate surfaces and inside flowfield, and then completing the integration over the related velocity phases. For diffusely reflective surfaces, the VDFs are related to the densities and temperatures at the two exits and the plate temperatures. For surfaces with specular reflections, the VDFs at the plate surface and inside the flowfield are identical, and independent of the surface temperature ratio and the geometric aspect ratio. Based on the VDFs and velocity phases, surface property coefficients (e.g., $C_p$, $C_f$, and $C_q$) and flowfield properties (e.g., density, velocity components, and temperature) are obtained. For the diffusely reflective surface scenario, the mass flow rate can be approximated and the results include four non-dimensional parameters: the aspect ratio, the density ratio, and two temperature ratios. For specularly reflective surface scenario, the surface and flowfield properties are uniform everywhere, the channel aspect ratio and plate temperatures do not have any influence. Particle simulations with the direct simulation Monte Carlo (DSMC) method are performed, and essentially identical results validate the theoretical work. This work is heuristic and can be used to investigate less rarefied micro-channel gaseous flows, for example, aid experimental measurement design for thermal transpiration flows. }

\begin{document}
\section{INTRODUCTION}
Micro-channels are important components widely used in MEMS and NEMS. Due to the tiny characteristic length, the interior gas flows could be highly rarefied or even free molecular. How free molecular gas flows inside a micro-channel is a very fundamental question, and there are numerous reports in the literature. For example, the mass flow rate is one important property and there are several factors that may have effects, e.g., the gas pressure and temperature differences between the two channel exits, the channel surfaces are diffusely or specularly reflective. Here, a diffuse surface means that when a particle hits the surface, it bounces off with a velocity of uniform probabilities within a specific solid angle. A specularly reflective surface means that it bounces off having the normal velocity component reversed but the tangent component unchanged. Realistic reflections are generally between these two limiting scenarios; hence, we can understand the micro-channel flows by studying them between plate surfaces with fully diffusely or  fully specular reflections. 

Because the problem is very fundamental, there are numerous reports in the literature, and we can only report a few as follows. 

The starting work on rarefied gas flows through channels of different cross-section was done by  Knudsen \cite{Knudsen}, Gaede \cite{Gaede}, Smoluchowski \cite{Smoluchowski} and Present \cite{Present}. Their work were both experimental and theoretical.   Later, Takao \cite{Takao} developed a new mathematical theory of rarefied gas flows.  Cercignani \cite{Cercignani1} used an asymptotic method and studied gas flows through a micro-channel with an inverse Kn number ranging from 0 to 10.5, i.e. the collisionless flow limit scenario is included. However, a constant pressure gradient through the channel was assumed in that study and it is not the situation of a long micro-channel.   Two-dimensional, highly rarefied gas flows with few collisions were also investigated with an isothermal assumption and linear pressure distribution \cite{Cercignani2}.  DeMarcus \cite{DeMarcus1, DeMarcus2, DeMarcus3} reported a systematic discussion on the free molecular flow problem for one-dimensional systems by using the Clausing's integral equation \cite{Clausing1, Clausing2} for the transmission probability of molecules crossing the channel. That method allows feasible iterations by hand. However, the results are limited to simple 1D situations and mainly include mass flow rate. The gas flows are assumed isothermal and collisionless. The work focused on a circular capillary, a pair of semi-infinite parallel plates and a highly dispersed bed of isotropic-ally scattered spheres. That work is classic but rather simple in computing detailed and complex 2D free molecular flowfield. Following DeMarcus's work, Berman \cite{Berman,Bermanerror} reported explicit expressions on the mass flow rates for collisionless flows through a capillary, flat plates, and beds of spheres. Their results indicate  that the transport in pressure volume units per unit time is proportional to the pressure difference, and obviously they adopted an averaged temperature which can not reflect the temperature differences between the exits or the plate surfaces. In 1972, Dayton \cite{Dayton} discussed molecular flows through ducts with sharp turns and bends.  In the same year, numerical and gaskinetic investigations \cite{Raghuraman} were reported on the study of highly dilute gas flows through a finite length slot. The results include an empirical formula for the mass flow rate, and it is quite worthy of reading.   In 1986, Santeler \cite{Santeler} provided new improvement on solving the Clasuing equation with several corrections.

Steckelmacher \cite{Steckelmacher} offered a review on the work related to high Kn number flows done before 1985. Most investigations on rarefied gas flows through a micro-channel focused on numerical simulations, e.g., the direct simulation Monte Carlo method \cite{bird, Moran, Roohi2, IP, IPCai, roohi}. Reese, Gallis and  Lockerby \cite{Reese} presented DSMC and super-hydrodynamic methods in studying non-equilibrium hypersonic gas flows and micro-device (e.g. micro-channels, micro-beams) flows. 

Cercignani \cite{Cercignani2004} provided a review on various methods to study micro-channel flows with various analysis work.  It is remarkable that Titareva and Shakhov \cite{Titareva} developed a linearized gaskinetic S-model to compute non-isothermal, rarefied gas flows inside a finite micro-channel, and reported that the mass flow rate is related to the temperature and pressure differences at the two exits, if the Kn number is low. The capillary end effect on micro-channel flow is also successfully investigated by using the gaskinetic method \cite{Sharipov}.

 Takata et al \cite{Takata1, Takata2} studied the problem of separating two different gas species by using micro-channels or micro-pumps with high Kn number, where the micro-pumps can be simplified as a planar channel with linear temperature distributions.  In 2009, Ye \& Yang \cite{ye} performed DSMC simulations to investigate the surface temperature effect on the heat and mass transfer. It is for situations with inter-molecular collisions. Shen \cite{cshen} discussed the Poiseuille alike flows inside micro-channels with pressure gradients. Both papers are interesting for reading, as work related to the current study. Livesey \cite{Livesey} presented work on pressure driven flows within long and short ducts of typical duct cross-sections, from the molecular to continuum flow limits. He developed a model, where the pressure difference is still an explicit factor. 
 

In 2011, Dongari, Zhang and Reese \cite{Reese2} proposed a power-law based  model so that the Navier-Stokes-Fourier equations can be employed for the transition-regime flows in typical of gas micro/nano-devices.  The model is derived for a system with planar wall confinement by taking into account the boundary limiting effects on the molecular free paths. It is applied to fully developed pressure-driven and thermal creep gas flows in micro-channels. More recently, the unified gas-kinetic  scheme (UGKS) is adopted to simulate \cite{Sliu} gas flows through micro/nano-channels. That method models the gas evolution with the gaskinetic method, both  elastic and inelastic collisions are taken into consideration. The simulated examples include thermal-conduction and pressure-driven flow within a wide range of Kn numbers.  Yuan and Rahman \cite{Duan} adopted a multiple-relaxation-time LBM with a second-order slip boundary condition to simulate highly rarefied micro-channel flows,  with the aid of a Bosanquet-type effective viscosity.  They reported that this method has several merits over other simulation methods. Based on the Boltzmann model equation,  a new gas-kinetic simulation scheme is used to study two-dimensional micro-scale gas flows with irregular configurations including pressure-driven micro-channel flows\cite{zhihui_Li}. The Kn  is within the range from  100 to 0.01. Srinivasan \cite{Srinivasan} et. al. performed numerical simulations and experimental validations on highly rarefied micro-channel gas flows. They adopted higher order slip boundary conditions, and used commercial software package Fluent which is more proper to simulate small Kn number gas flows.

The current work is motivated by several formulas for mass flow rate  for collisionless gas flows through a micro-channel.  For the flat plate scenario, the following formula was suggested by Berman \cite{Berman,Bermanerror}:
\begin{equation} 
G =  \frac{1}{4} Av \Delta p  \left( \frac{1}{2} (1+ \sqrt{1+L'^2} -L' ) - \frac{ \frac{3}{2}(L'-\mbox{ln}(L'+ \sqrt{L'^2+1}) )^2}{L'^3+3L'^2 +4 -(L'^2 +4) \sqrt{1+L'^2}} \right),
\label{eqn:berman}
\end{equation}
where $G$ is the transport in pressure volume units per unit time, $A$ is the cross-section area, $v$ is the mean thermal speed of the gas, $\Delta p$ is the pressure difference, $L'$ is the planar channel aspect ratio of length to height. Evidently, there is no temperature ratio related to the mass flow rate. Similar limitation can be found in the formulas suggested by Cercignani \cite{Cercignani2004}. Shall we include one or several temperature ratios for general collisionless flow cases? Shall the results be different for the diffusely and specularly reflective planar channel cases? Shall the pressure difference be one crucial factor?

This paper aims to address this fundamental problem of how  collisionless gas flows inside a planar micro-channel, especially the above questions related to the mass flow rate. We use the gaskinetic theory and develop analytical solutions for the flowfield and surface properties, and identify important factors, including several temperature ratios.

The rest of this paper is organized as follows. Section \ref{sec:diffuse} discusses the situation with diffusely reflective channel surfaces; Section \ref{sec:specular} discusses the situation with specularly reflective surfaces; and Section \ref{sec:conclusions} summarizes this study with several conclusions.

\section{CHANNELS WITH DIFFUSELY REFLECTIVE SURFACES}
\label{sec:diffuse}
It is assumed that free molecular gas flows through a planar micro-channel with a length $L$ and a height $H$. The gas at the left entrance or exit has a number density $n_L$ and a temperature $T_L$. Correspondingly, the values for the right entrance are $n_R$ and $T_R$. The temperatures for the top and bottom plate surfaces are  $T_T$ and $T_B$. This section discusses the situation that surfaces have diffuse reflections. 
\begin{figure}[ht]
\centering
\includegraphics[trim=0 110 0 100, clip, width=4.5in, height=2.3in]{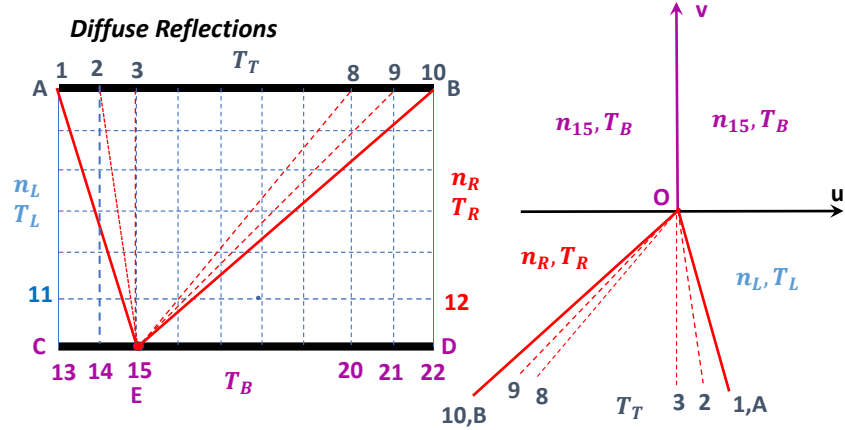}
\caption{Velocity phase construction for Point $E$ at the bottom channel surface. Left: problem illustrations. Right: the corresponding velocity phase for Point $E$. }
\label{fig:illustration_p1}
\end{figure}

The approach adopted in this work follows fundamental gaskinetic theory, and the method can be considered as extensions of our past work on internal and external free molecular flows \cite{enclosure, cai_jsr, aiaaj_cai}.

Figure \ref{fig:illustration_p1} illustrates how to construct the velocity phases for a surface point, and this construction is important for the surface and flowfield property computations. For convenience of discussions, both the top and bottom plates are divided into 10 segments, but in the real computations, much more segments have been used. The right side of this figure shows the corresponding velocity phase for Point $E$ at the bottom surface. The Velocity Distribution Functions (VDFs) for the gas entering from the left and right entrances are assumed to be Maxwellian due to free molecular flow. Another two Maxwellian VDFs are used to model those molecules bouncing off the top and bottom plates for the diffusely reflective surfaces: 
\begin{equation}
    f_i(u,v) = n_i(x)  \frac{\beta_i}{\pi}  e^{- \beta_i (u^2+v^2)},
\label{eqn:surfvdf}
\end{equation}
where $\beta_i = 1/(2RT_T)$ for the top plate and $1/(2RT_B)$ for the bottom plate; $R$ is the specific gas constant; the number densities in these two functions vary with the local positions on the plate and need to be determined as the first step. For convenience, $n_i(x)$ is defined as a ``virtual'' density. 

By using the Maxwellian VDF and velocity phases for Point $E$ at the bottom plate as an example, as shown in Fig. \ref{fig:illustration_p1}, the surface properties can be derived as follows by using the gaskinetic theory (e.g., by Kogan \cite{Kogan}). 

The virtual number density in Eqn. \ref{eqn:surfvdf} can be determined by using the condition of zero flux at the plate surface, or the non-penetration condition. At the bottom surface Point $E$, there are three groups of incoming molecules, from the left and right entrances, and the top plate, respectively. There is also a group of outgoing molecules bouncing off point $E$ with a VDF described by Eqn. \ref{eqn:surfvdf}. According to the zero-flux condition, the virtual density $n_E$ can be determined as: 
\begin{equation}
\begin{array}{rll}
  \frac{n_E(x)}{n_L} &=&  \frac{1}{n_L} \left(                            \int_{\Omega_{AOu} } f_L dudv + 
                       \int_{\Omega_{BO(-u)}} f_R dudv + 
                       \sum_{i=1}^9\int_{\Omega_{iO(i+1)}} f_{T,i} dudv  \right) \\
                     &=& \frac{1}{2} \left( \sqrt{ \frac{T_L}{T_B}} ( 1- \cos \alpha_L)   + \frac{n_R}{n_L}  \sqrt{ \frac{T_R}{T_B}} (1- \cos \alpha_R)  +
  \sqrt{\frac{T_T}{T_B}} \sum_{i=1}^{9} \frac{n_i +n_{i+1}}{2n_L}  ( \cos  \alpha_i -  \cos \alpha_{i+1} )   \right),
\end{array}
\label{eqn:nbottom}
\end{equation}
where $\alpha_L = \angle AEC = \angle AOu$ and $\alpha_R = \angle BED = \angle BO(-u)$, they are the acute angles subtended by the left and right exits, respectively, and $\alpha_i$ is the angle $ 2 \pi-\angle iEC $ shown in the left side of Fig. \ref{fig:illustration_p1} or $ 2 \pi-\angle iOu $ shown in the right side of Fig. \ref{fig:illustration_p1}. The transformation from the Cartesian coordinate to the polar coordinate for the integral in the velocity phase is used, which simplifies the above derivations thanks to the  symmetric property of VDFs with respect to the origin. 

\begin{figure}[ht]
    \begin{minipage}[l]{0.48\textwidth}
      \centering
      \includegraphics[width=3.8in]{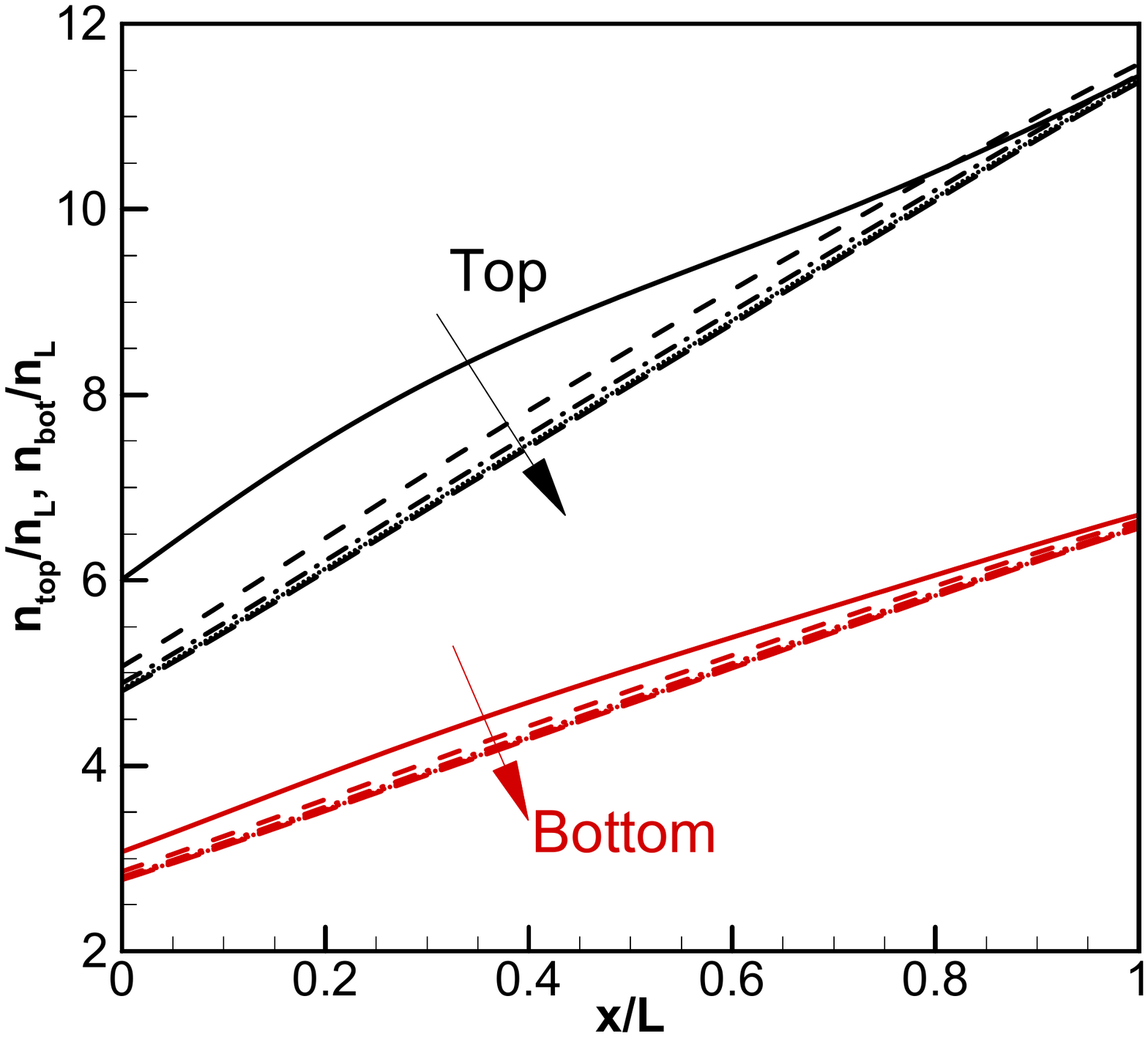}
  \end{minipage}
  \begin{minipage}[l]{0.48\textwidth}
     \centering
      \includegraphics[width=3.8in]{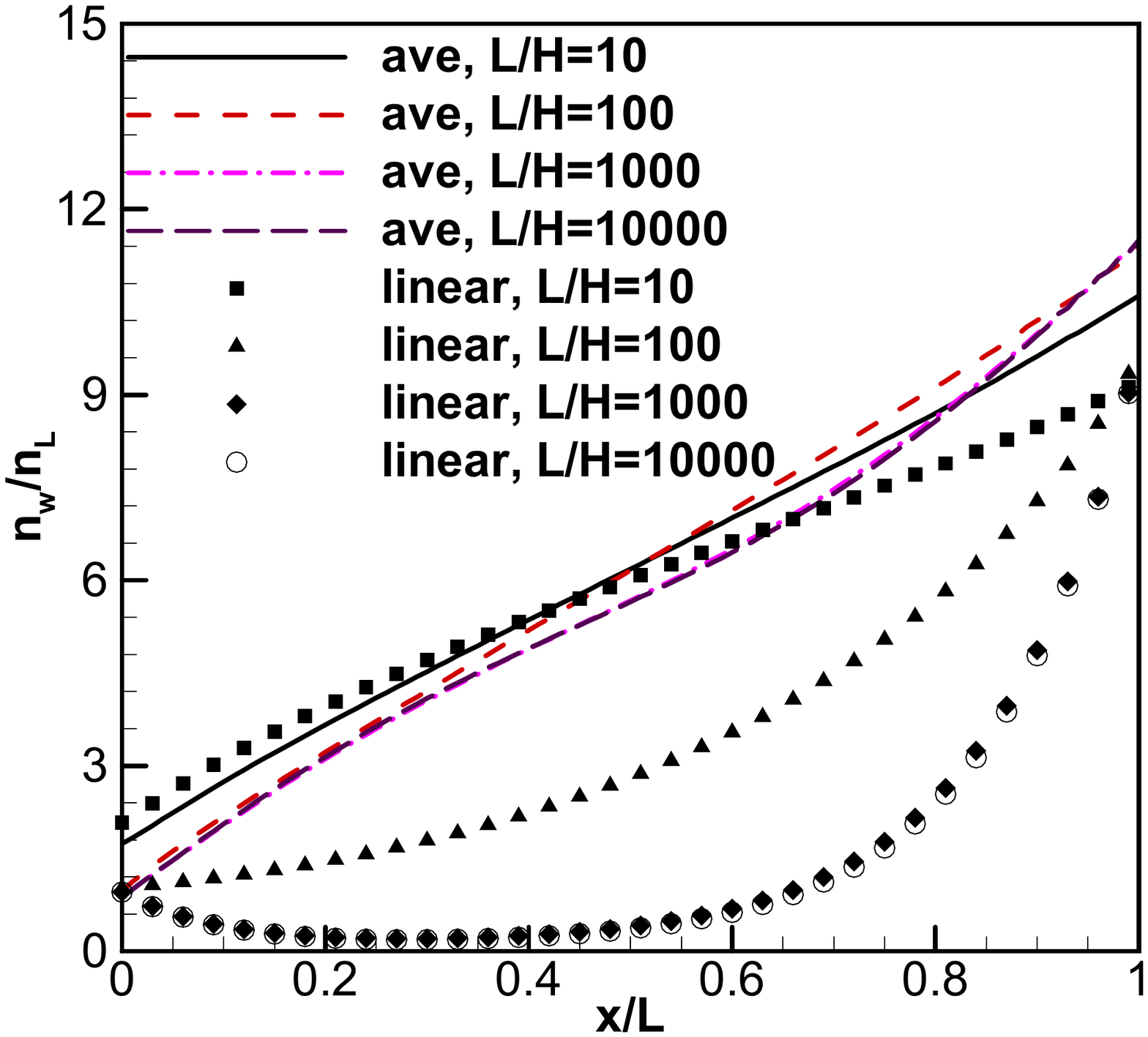}
  \end{minipage}
\caption{Left: virtual number density profiles along the top (black) and bottom (red) plate surfaces, $T_L:T_R:T_T:T_B=4:2:1:3$, $n_L:n_R=1:10$, and $L/H=2.$  Right: virtual number density profiles with different aspect ratios, $L/H=10$, 100, 1000, and 10000;  $n_L:n_R=1:10$, $T_R:T_L =2:1$, the surface temperatures are either constant with $T_T=T_B = (T_L+T_R)/2$, or linear, $T_T(x)=T_B(x) = T_L + (x/L) (T_R-T_L).$}
\label{Fig:nt_nb}
\end{figure}

The above virtual density distribution is crucial and it includes three temperature ratios, a density ratio, and a geometry ratio ( embedded in $\alpha_L$ and $\alpha_R$). It is understandable that the virtual number density distribution along one plate surface affects that along the other surface, i.e., $n_E$ at the bottom surface depends on $n_i$ at the top surface, and vice versa. These two distributions can be determined by using iterations. Figure \ref{Fig:nt_nb} (left) shows the developments with 5 iterations of the normalized virtual density distributions. At each point, non-penetration conditions must be satisfied. To clearly demonstrate the effects from various factors, the four temperatures are assigned with different values, and the channel is short with $L/H=2.0$. Each plate surface is divided into 100 short segments. The initial virtual density distributions are set to zero along the two plates, and they develop and converge fast to the two final linear distributions. Numerical solutions indicate that with the same plate surface temperatures, these two virtual density distributions merge into one, which is expected and thus omitted here.

The right side of Fig. \ref{Fig:nt_nb} illustrates the surface virtual number density distributions with two different surface temperature profiles. The top and bottom plates have the same surface temperature here and can be simply set as constant as the average temperature $T_T=T_B=(T_L +T_R)/2$, or it can be assumed more reasonably with a linear distribution from $T_L$ to $T_R$.  The latter may happen for thermal transpiration micro-channel flows with a large $L/H$ ratio. The right figure illustrates that if a linear temperature profile is assumed and the channel aspect ratio is large, e.g., $L/H >100$, the virtual number density profiles at the surface become highly non-linear. This  phenomenon may give  insights for modelling thermal transpiration flows\cite{transpiration}. Whether the flows are choked inside such a long channel is beyond the scope of this work. 

For a diffusely reflective surface, the interesting surface properties may include slip velocities, pressure, friction and heat flux. The surface slip velocity is mainly due to the three groups of incoming molecules, and these outgoing or diffusely bouncing-off molecules contribute less to the slip velocity at this local point. This is because the averaged surface tangent velocity for the last group is zero,  although the mass carried by this group needs to be included in computing the slip velocity as the overall average. By using Fig. 
 \ref{fig:illustration_p1} (right) as an example, the local slip velocity at point $E$ is:
\begin{equation}
\begin{array}{rll}
  \frac{u_{slip,E}}{ \sqrt{2RT_L} }  &=&  
    \frac{1}{n_E/2 + \Lambda_B + \Lambda_L+ \Lambda_R} ( \int_{\Omega_{AOu} } u f_L dudv + 
                 \int_{\Omega_{BO(-u)}} u f_R dudv + 
                 \sum_{i=1}^9\int_{\Omega_{iO(i+1)}} u f_{T,i} dudv  )  \\
        &=&\frac{n_L}{n_E/2 + \Lambda_B + \Lambda_R+ \Lambda_L} \frac{1}{ 4 \sqrt{\pi} } \left(  \sin \alpha_L -  \sqrt{\frac{T_R}{T_L}} \frac{n_R}{n_L} \sin \alpha_R   +  \sqrt{\frac{T_T}{T_L}} \Lambda_T \right),
\end{array}
\end{equation}
\begin{equation*}
\begin{array}{rll}
 &\Lambda_B = \sum_{i=1}^{9}  \frac{\alpha_{i+1} -\alpha_i}{2 \pi} \frac{n_i 
 + n_{i+1}}{2},&     \Lambda_T = \sum_{i=1}^{9}  \frac{n_i+n_{i+1} }{2 n_L}  (\sin \alpha_{i+1}  - \sin \alpha_i ),  \alpha_i = 2 \pi - \mbox{acos}  \frac{x_E-x_i}{ \sqrt{ (x_i -x_E)^2 +H^2 } },  \\
 &\Lambda_L =  \frac{\alpha_L}{2 \pi} n_L, & \Lambda_R= \frac{\alpha_R}{2 \pi} n_R,
\end{array}
\end{equation*} 
where $\Lambda_B$ and $\Lambda_T$ represent the contribution from the whole top plate;  $\Lambda_L$ and $\Lambda_R$ represent contribution from the left and right entrances; $n_E$ represents the virtual number density of bouncing-off molecules at Point $E$, ``$1/2$'' is because only half of those molecules are moving up.

The local surface pressure coefficient, $C_p$, along the surfaces is derived as: 
\begin{equation}
\begin{array}{rll}
    C_{p,E} &=&   
        \frac{1}{n_LkT_L}   \left( \frac{n_E kT_B}{2}  +      \int_{\Omega_{AOu} } v^2 f_L dudv + 
                 \int_{\Omega_{BO(-u)}} v^2 f_R dudv + 
                 \sum_{i=1}^9\int_{\Omega_{iO(i+1)}} v^2 f_{T,i} dudv  \right)      \\
            &=&\frac{1}{2} \frac{n_E }{n_L} \frac{T_B}{T_L} + \frac{1}{2 \pi} \left( \alpha_L - \frac{1}{2} \sin 2 \alpha_L \right) +  \frac{1}{2 \pi} \frac{T_R}{T_L} \frac{n_R}{n_L} \left( \alpha_R - \frac{1}{2} \sin 2 \alpha_R \right)  +  \\
        &&\frac{1}{2 \pi} \frac{T_T}{T_L} \sum_{i=1}^{9} \frac{n_{i} +n_{i+1}}{2n_L}  \left(  \alpha_{i+1} - \alpha_i - \frac{1}{2} \sin 2 \alpha_{i+1} + \frac{1}{2} \sin2 \alpha_i   \right).
\end{array}
\label{eqn:Cp}
\end{equation}
Those particles bouncing off the diffuse plate surface do not contribute to the local surface friction force due to zero average tangential velocity/momentum, but do affect the local surface heat flux as their average kinetic energy is nonzero: 
\begin{equation}
\begin{array}{rll}
 C_{f,E} &=& \frac{1}{n_LkT_L}   \left( \int_{\Omega_{AOu} } uv f_L dudv + 
    \int_{\Omega_{BO(-u)}} uv f_R dudv+
    \sum_{i=1}^9\int_{\Omega_{iO(i+1)}} uv f_{T,i} dudv  \right)  
 \\
    &=& \frac{1}{2\pi} \sin^2 \alpha_L  - \frac{1}{2 \pi} \frac{n_R}{n_L} \frac{T_R}{T_L}   \sin^2 \alpha_R +  
        \frac{1}{4 \pi} \frac{T_T}{T_L} \sum_{i=1}^{9} \frac{n_i +n_{i+1}}{n_L}  \left( \sin^2 \alpha_{i+1} - \sin^2 \alpha_i   \right), 
\end{array}
\label{eqn:Cf}
\end{equation}
\begin{equation}
\begin{array}{rll}
    C_{q,E} &=&   \frac{1}{n_L k T_L \sqrt{2RT_L}} 
       \left(-\int_{v>0 } e f_E  + \int_{\Omega_{AOu} } e f_L  + \int_{\Omega_{BO(-u)}} e f_R  + 
    \sum_{i=1}^9\int_{\Omega_{iO(i+1)}} e f_{T,i}
       \right)dudvdw  \\
      &=&  \frac{1}{2\sqrt{\pi} }  \bigg[  1- \cos \alpha_L   + \frac{n_R}{n_L} {\sqrt{ \frac{T_R}{T_L } }}^3 (1 - \cos \alpha_R)  - 2 \frac{n_E }{n_L}  {\sqrt{ \frac{T_B}{T_L } }}^3   \\
        &&   + {\sqrt{ \frac{T_T}{T_L } }}^3 \sum_{i=1}^{9} \frac{n_i +n_{i+1}}{2n_L}  \left( \cos \alpha_i - \cos \alpha_{i+1} \right)  \bigg].
\end{array}
\label{eqn:Cq}
\end{equation}
where $e=v [ u^2+v^2+w^2]/2$.

To validate the above surface properties, several simulations are performed with the DSMC method. A well-tested DSMC simulation package \cite{GRASP} is adopted for the simulations. The simulation domain is divided into a mesh with $100\times100$ points. 

\begin{figure}[ht]
 \begin{minipage}[l]{0.48\textwidth}
      \centering
      \includegraphics[width=3.8in]{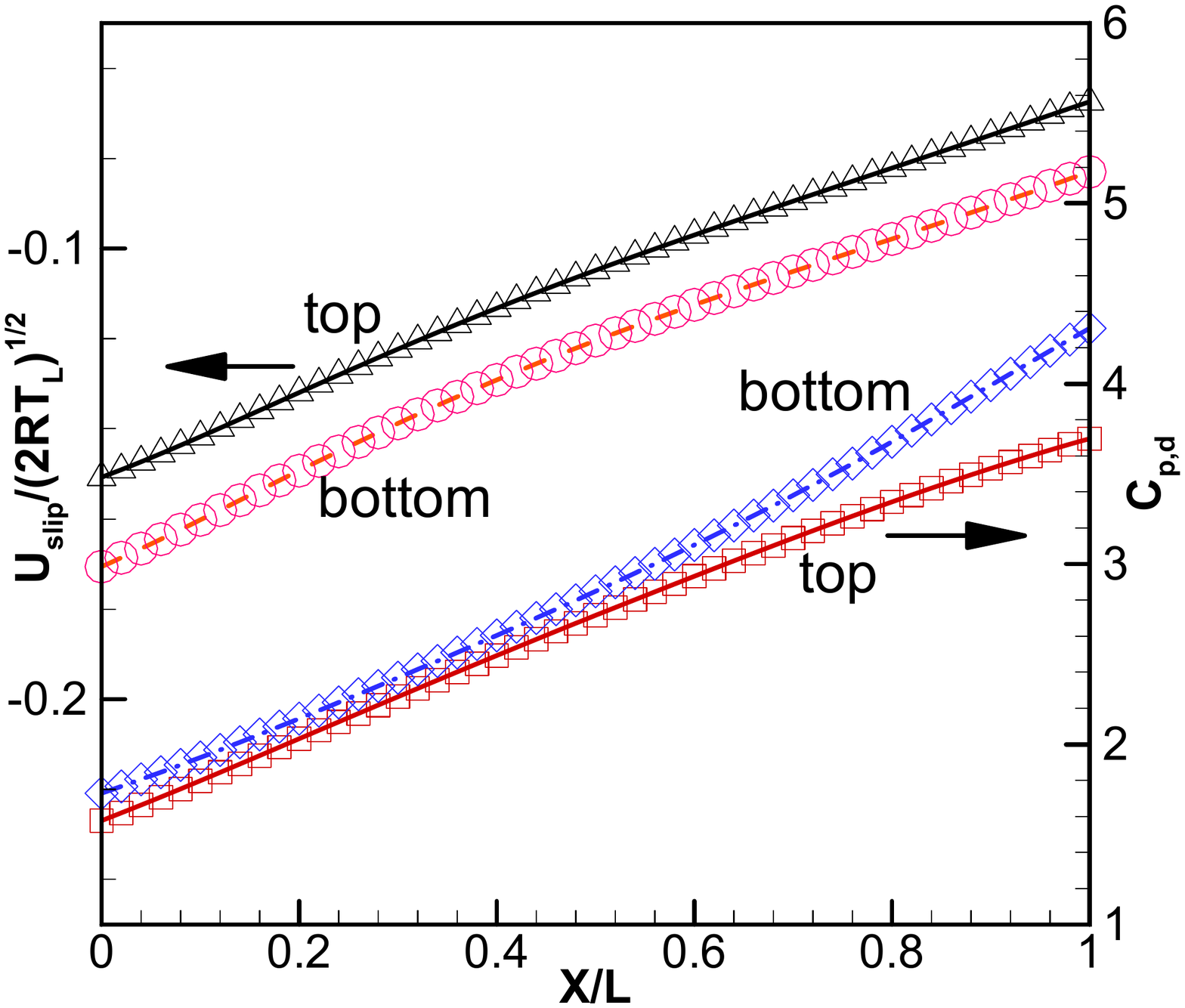}
  \end{minipage}
  \begin{minipage}[l]{0.48\textwidth}
     \centering
      \includegraphics[width=3.8in]{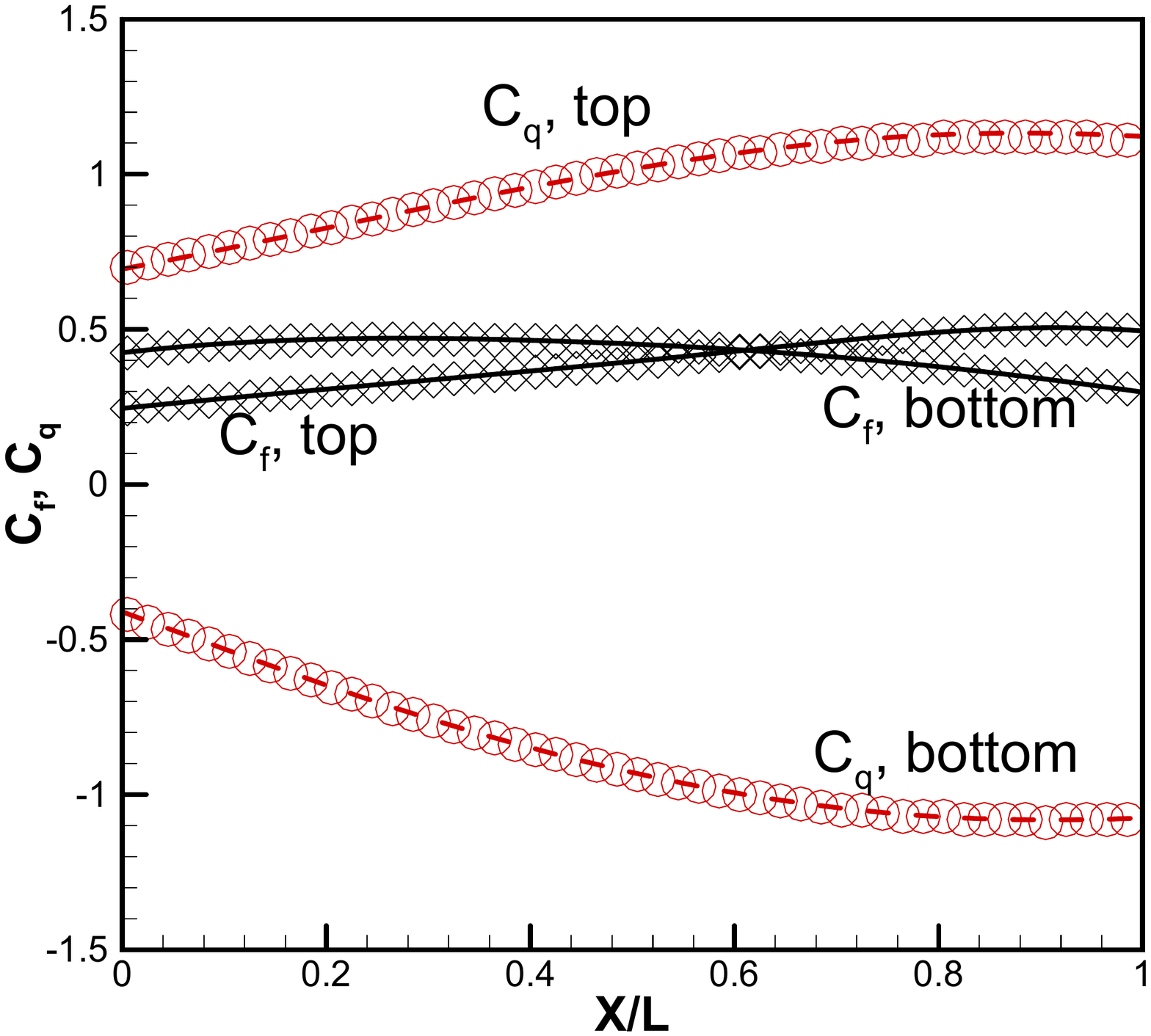}
  \end{minipage}
\caption{Left: diffuse surface slip velocities, $u_{slip}/\sqrt{2RT_L}$, and pressure coefficients $C_p$ along the top and bottom surfaces. Right: the corresponding coefficients for surface friction $C_f$ and heat flux $C_q$. $T_L:T_R:T_T:T_B=4:2:1:3$, $n_L:n_R=1:10$, and $L/H=2.$  Symbols: DSMC; Lines: analytical.}
\label{Fig:slip_cpd_cf_cq} 
\end{figure}

Figure \ref{Fig:slip_cpd_cf_cq} (left) presents typical surface slip velocities and pressure coefficients along the top and bottom surfaces. The DSMC simulation results are plotted as symbols and the analytical results are plotted with lines. To clearly demonstrate the differences between the two plates, the surface temperatures $T_T$ and $T_B$ are different and the channel aspect ratio is $L/H=2$. Evidently, these profiles are nonlinear. The pressure closer to the right exit is relatively larger, but the velocity closer to the left exit is relatively larger (the negative value means that the flow is from the right to the left). It seems that this is a typical pressure driven free molecular flow with a pressure difference of $n_R kT_R -n_L k T_L$.  Figure \ref{Fig:slip_cpd_cf_cq} (right) shows the friction and heat flux profiles along the two plate surfaces. The nonlinear and unsymmetrical patterns are evident. The simulation results and the above analytical solutions are essentially identical. 
\begin{figure}[ht]
\centering
\includegraphics[trim=0 120 0 100, clip, width=4.5in, height=2.3in]{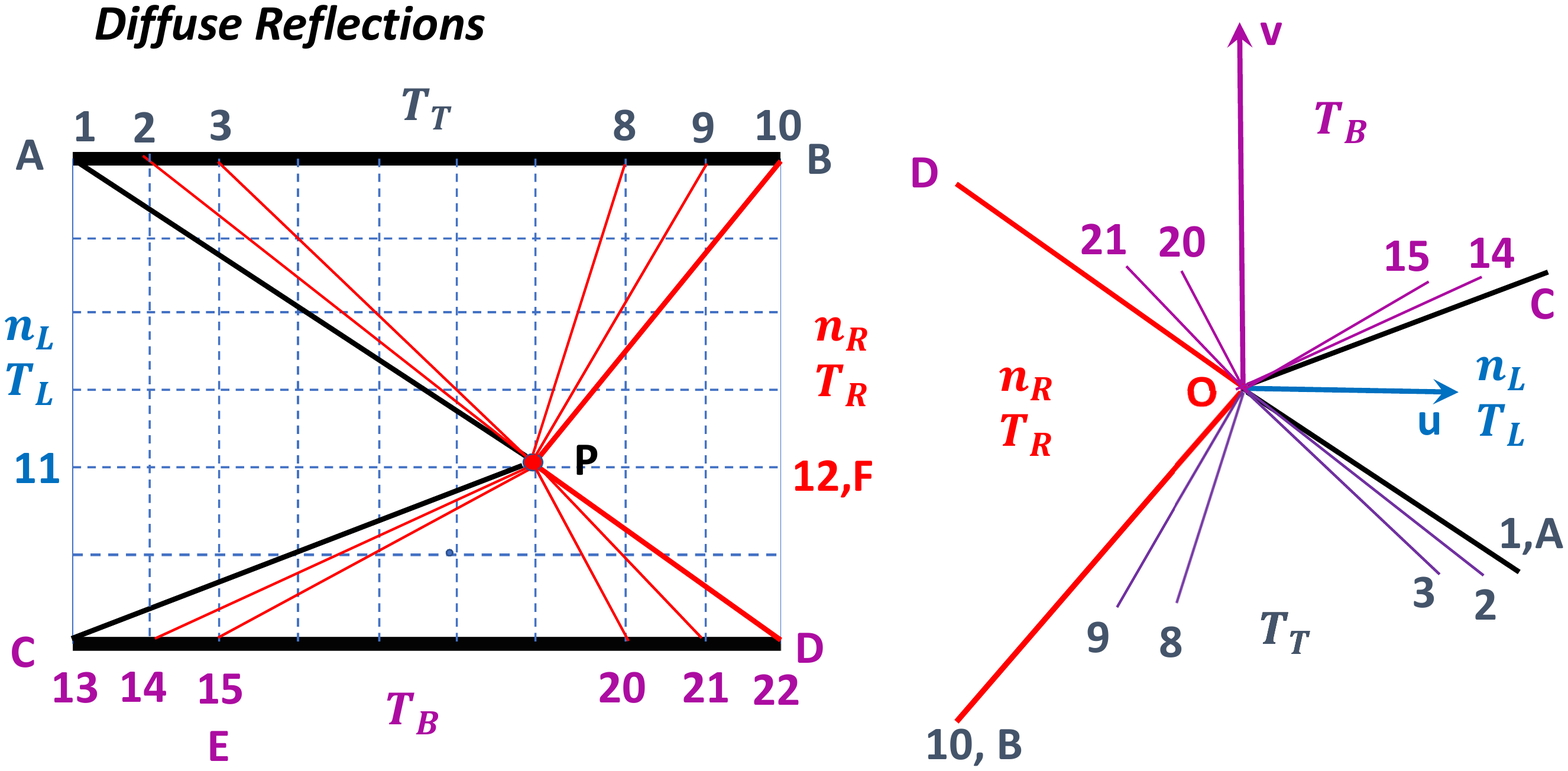}
\caption{Left: flowfield property computations at point $P(X,Y)$; Right: the corresponding velocity phase for point $P$.}
\label{fig:illustration_flowfield}
\end{figure}

Figure \ref{fig:illustration_flowfield} sketches a flowfield point $P(X,Y)$ and the related velocity phase for that point. The computations of related flowfield properties include contributions from the two exits, the top and bottom plates. The formula for number density is:  
\begin{equation}
\frac{n(X,Y)}{n_L} = \frac{\alpha_{L1} + \alpha_{L2}}{2 \pi} + \frac{\alpha_{R3} + \alpha_{R4}}{2 \pi} \frac{n_R}{n_L} + \frac{1}{ 2 \pi} \sum_{i=1}^{9} \frac{n_i +n_{i+1} }{2n_L} \alpha_i + \frac{1}{ 2 \pi} \sum_{i=13}^{21} \frac{n_i + n_{i+1} }{2n_L} \alpha_i,  
\label{Eqn:nxy}
\end{equation}
where $\alpha_{L1}= \angle AOu$, $\alpha_{L2} = \angle COu$,  $\alpha_{R3} = \angle BO(-u)$, and $\alpha_{R4} = \angle DO(-u)$, as shown in Fig. \ref{fig:illustration_flowfield} (right); $\alpha_i$ here is the acute angle subtended by the corresponding segment at the top or bottom surfaces from the flowfield point $P(X,Y)$. The other flowfield properties are derived and the final expressions are: 
\begin{equation}
\begin{array}{rll}
\frac{U(X,Y)}{\sqrt{2RT_L}} &=& \frac{1}{4 \sqrt{\pi}} \frac{n_L}{n} \bigg[  
  \sin \alpha_{L1}  -  \frac{n_R}{n_L}  \sqrt{ \frac{T_R}{T_L}}  (\sin \alpha_{R3} + \sin \alpha_{R4} ) +  \sqrt{ \frac{T_T}{T_L} }   \sum_{i=1}^9  \frac{n_i + n_{i+1}}{2n_L}  (\sin \alpha_{i+1} - \sin \alpha_i)   \\
  && + \sin \alpha_{L2}  +\sqrt{ \frac{T_B}{T_L} }   \sum_{i=13}^{21}  \frac{n_i + n_{i+1}}{2n_L}  (\sin \alpha_{i+1} - \sin \alpha_i)     
\bigg],
\end{array}
\label{Eqn:Uxy}
\end{equation}
\begin{equation}
\begin{array}{rll}
\frac{V(X,Y)}{\sqrt{2RT_L}} &=& \frac{1}{4 \sqrt{\pi}} \frac{n_L}{n} \bigg[  
  \cos \alpha_{L2}  -  \frac{n_R}{n_L}  \sqrt{ \frac{T_R}{T_L}}  (\cos \alpha_{R4} + \cos \alpha_{R3} ) -  \sqrt{ \frac{T_T}{T_L} }   \sum_{i=1}^9  \frac{n_i + n_{i+1}}{2n_L}  (\cos \alpha_{i} - \cos \alpha_{i+1})   \\
  &&  - \cos \alpha_{L1}  +\sqrt{ \frac{T_B}{T_L} }   \sum_{i=13}^{21}  \frac{n_i + n_{i+1}}{2n_L}  (\cos \alpha_{i} - \cos \alpha_{i+1} )     
\bigg],
\end{array}
\label{Eqn:Vxy}
\end{equation}
\begin{equation}
\frac{T(X,Y)}{ T_L } =  \frac{1}{2 \pi} \frac{n_L}{n} \left[  \alpha_{L2} -\alpha_{L1}  + (\alpha_{R4} - \alpha_{R3} ) \frac{T_R}{T_L} \frac{n_R}{n_L}  + ( \frac{T_T}{T_L} \sum_{i=1}^9  +  \frac{T_B}{T_L} \sum_{i=13}^{21}  )    \frac{n_i+n_{i+1}}{2n_L} \alpha_i    \right] - \frac{2(U^2 + V^2) }{3RT_L},
\label{Eqn:Txy}
\end{equation}
\begin{equation}
\begin{array}{rll}
\frac{T_x(X,Y)}{ T_L } & = & - \frac{U^2} {RT_L}  + \frac{1}{2 \pi} \frac{n_L}{n}  \bigg[    \alpha_{L2} - \alpha_{L1} +\frac{\sin 2\alpha_{L2} - \sin 2 \alpha_{L1} }{2} + \frac{T_R}{T_L} \frac{n_R}{n_L} ( \alpha_{R3} - \alpha_{R4}  +\frac{\sin 2\alpha_{R3} - \sin 2 \alpha_{R4} }{2} )  +  \\
 & &  ( \frac{T_T}{T_L} \sum_{i=1}^9 + \frac{T_B}{T_L} \sum_{i=13}^{21} )  \frac{n_i+n_{i+1}}{n_L}        (\alpha_{i+1}-\alpha_i  + \frac{ \sin 2\alpha_{i+1} - \sin 2\alpha_i }{2}  )    \bigg],
\end{array}
\label{Eqn:Txxy}
\end{equation}
\begin{equation}
\begin{array}{rll}
\frac{T_y(X,Y)}{ T_L } & = & - \frac{V^2} {RT_L}  + \frac{1}{2 \pi}\frac{n_L}{n}  \bigg[   \alpha_{L2}-\alpha_{L1} -\frac{\sin 2\alpha_{L2} - \sin 2 \alpha_{L1} }{2} + \frac{T_R}{T_L} \frac{n_R}{n_L} (\alpha_{R3} -\alpha_{R4} -\frac{\sin 2\alpha_{R3} - \sin 2 \alpha_{R4} }{2} )  +  \\
 & & (\frac{T_T}{T_L} \sum_{i=1}^9  + \frac{T_B}{T_L} \sum_{i=13}^{21} )  \frac{n_i+ n_{i+1} }{2n_L} \left( \alpha_{i+1} - \alpha_i  - \frac{1}{2} ( \sin \alpha_{i+1} - \sin \alpha_i ) \right)   \bigg].
\end{array}
\label{Eqn:Tyxy}
\end{equation}

\begin{figure}[ht]
  \begin{minipage}[l]{0.48\textwidth}
      \centering
      \includegraphics[width=3.8in]{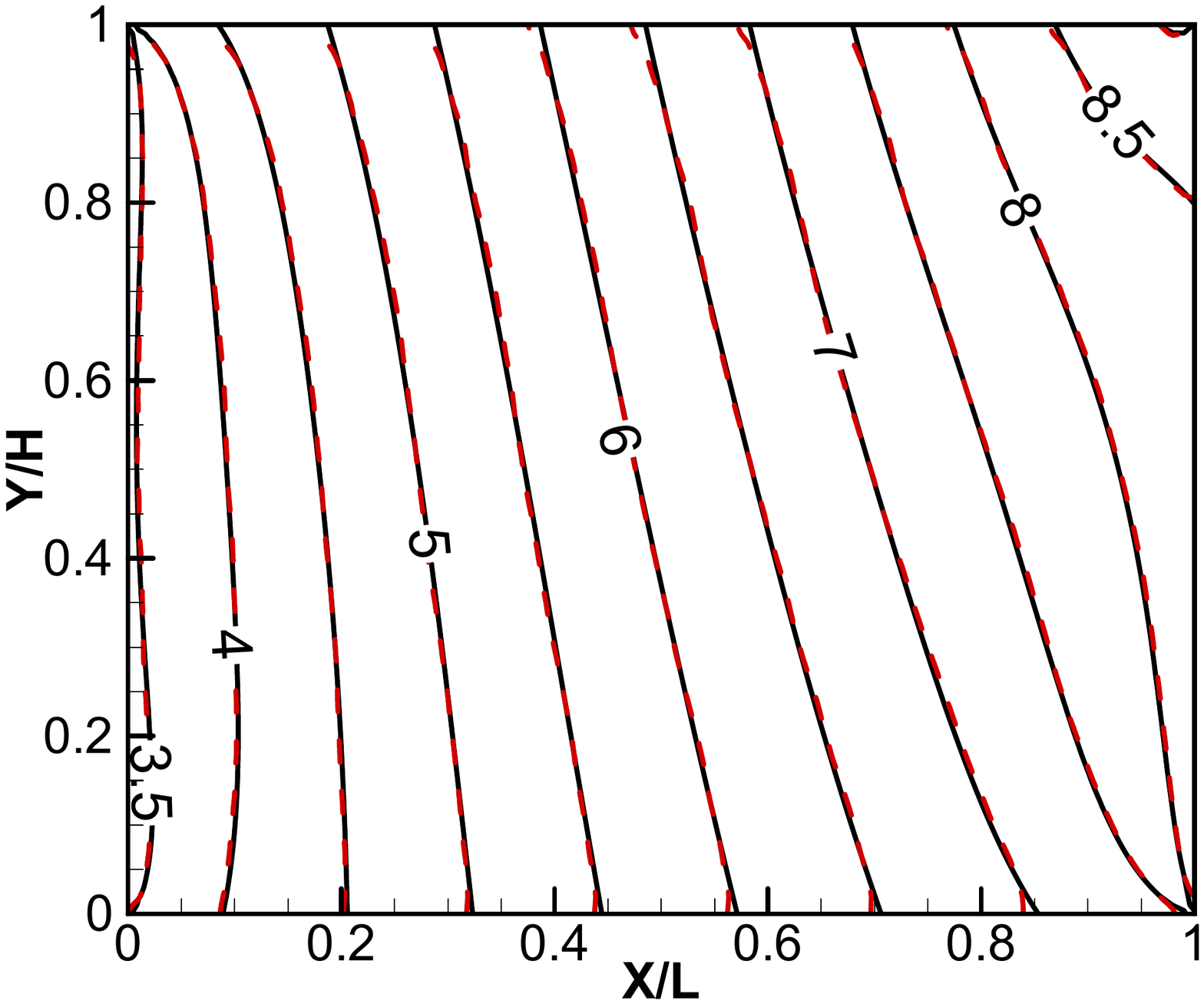}
  \end{minipage}
  \begin{minipage}[l]{0.48\textwidth}
      \centering
      \includegraphics[width=3.8in]{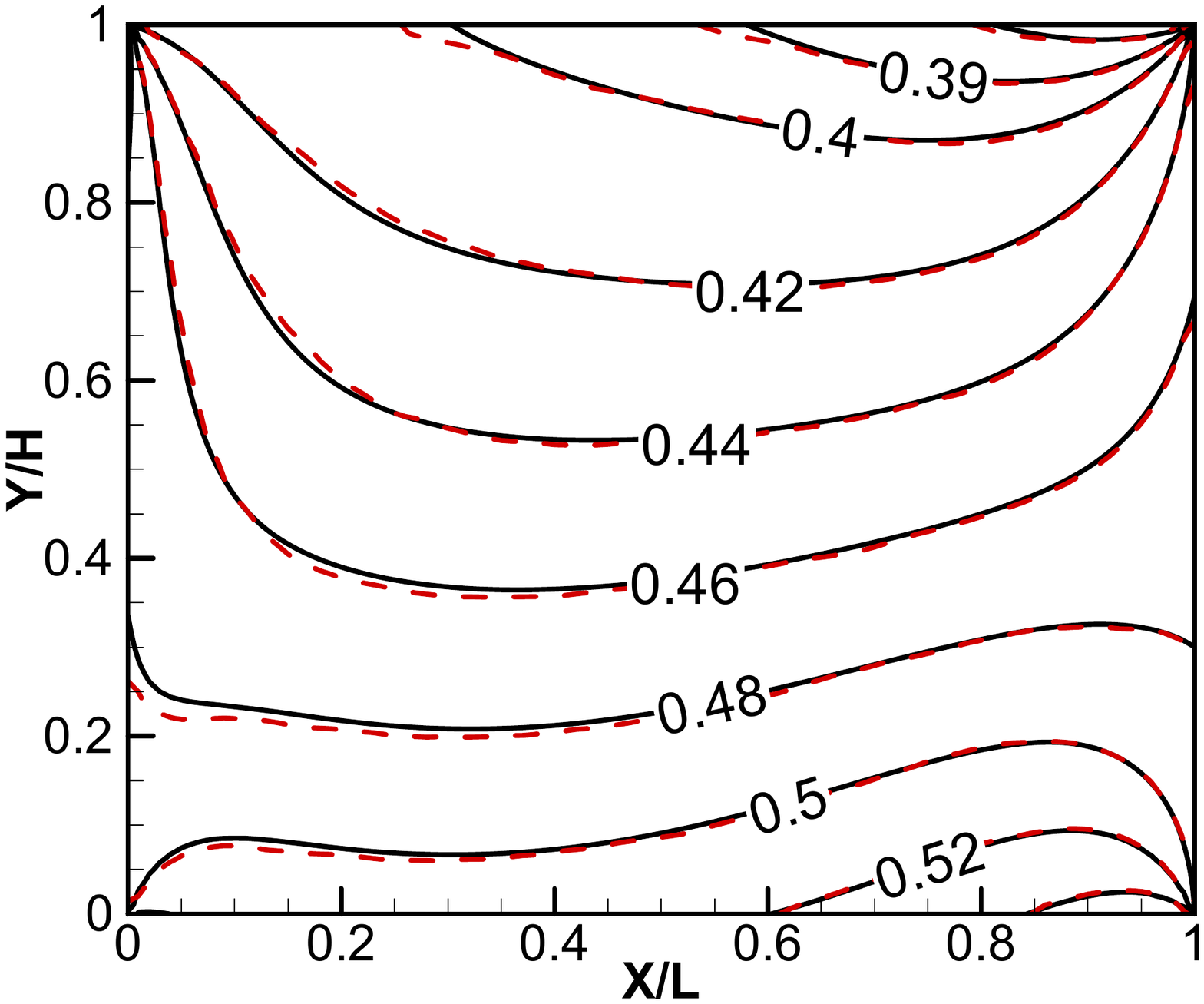}
  \end{minipage}
\caption{Left: flowfield density $n(X,Y)/n_L$;  Right: temperature $T(X,Y)/T_L$, with diffusely reflective surfaces. $T_L:T_R:T_T:T_B=4:2:1:3$, $n_L:n_R=1:10$, and $L/H=2.$ Dashed lines: DSMC; solid lines: analytical.}
\label{Fig:diffuse_nT}
\end{figure}

Figure \ref{Fig:diffuse_nT} shows the DSMC simulation results and analytical solutions for the flowfield distributions of number density and average temperature. It shall be reminded that this is a short channel, $L/H=2.0$, and  the strong variations in the flowfield are effective to validate the analytical solutions. If the aspect ratio $L/H$ is large, the variations become relatively mild and the agreements will be even better. Because $T_T$ and $T_B$ are different, the flowfield patterns are not symmetric. There are abrupt density changes around the exit corners. For example, for this test case with $n_R/n_L =10$, the density ratio at the left exit is close to 3.5, instead of 1.0. The density ratio at the right exit is close to 8.5, rather than 10. The density ratio at the middle station is close to 6.0, which is slightly larger than 5.5 or $(n_R+n_L)/(2n_L)$. They are because a short channel can allow molecules to cross it relatively easier. The relatively cold plate surfaces compared to the entrances may also affect the density distribution, as clearly shown in Eqn. \ref{eqn:nbottom}. The four corners with abrupt variations can be considered as singularity points. The analytical solutions are complex and quite challenging for validations. However, as shown clearly, the simulation and analytical solutions have very good agreements, showing that the final formulas are correct. 

\begin{figure}[ht]
  \begin{minipage}[l]{0.48\textwidth}
      \centering
      \includegraphics[width=3.9in]{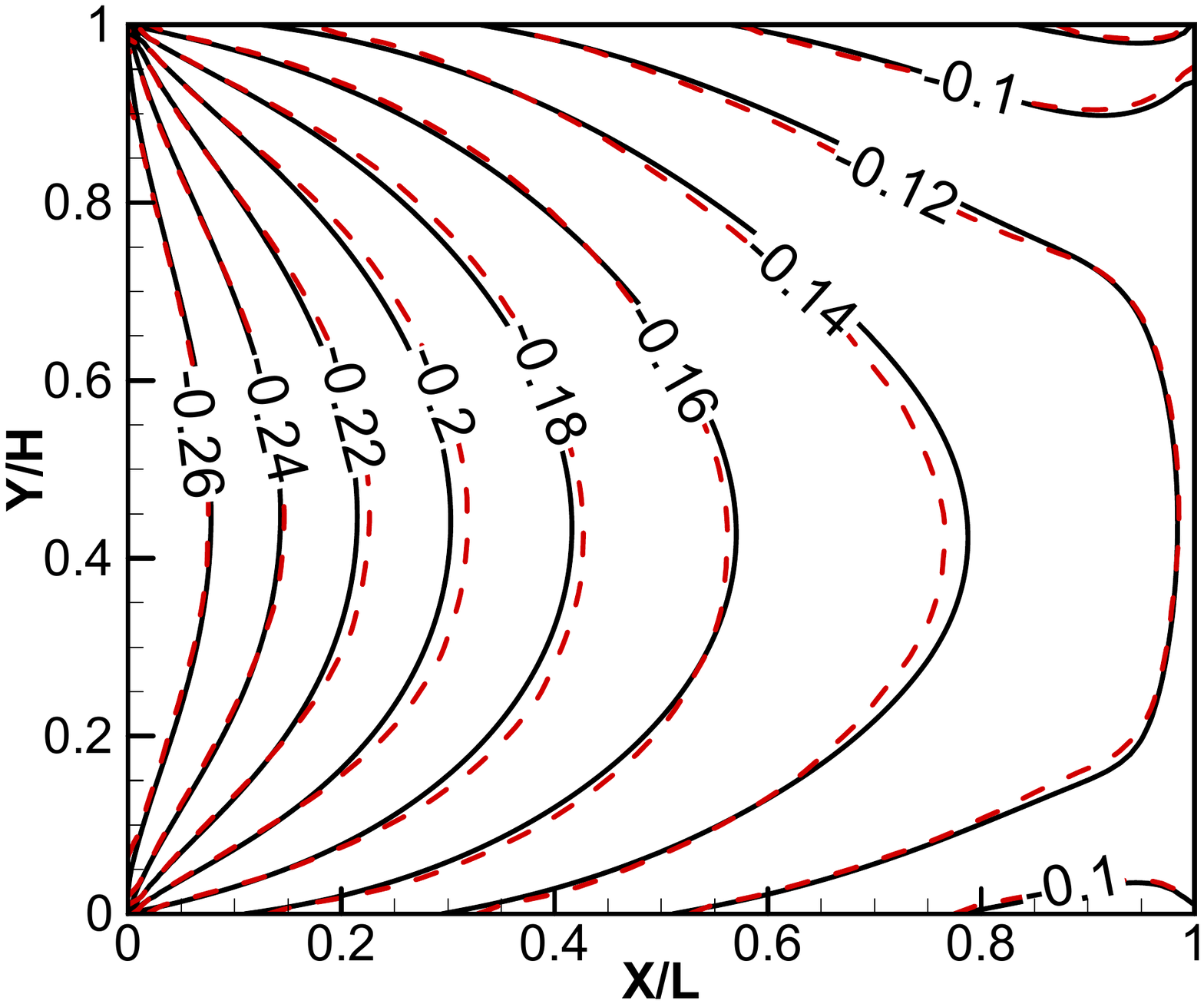}
  \end{minipage}
  \begin{minipage}[l]{0.48\textwidth}
     \centering
      \includegraphics[width=3.8in]{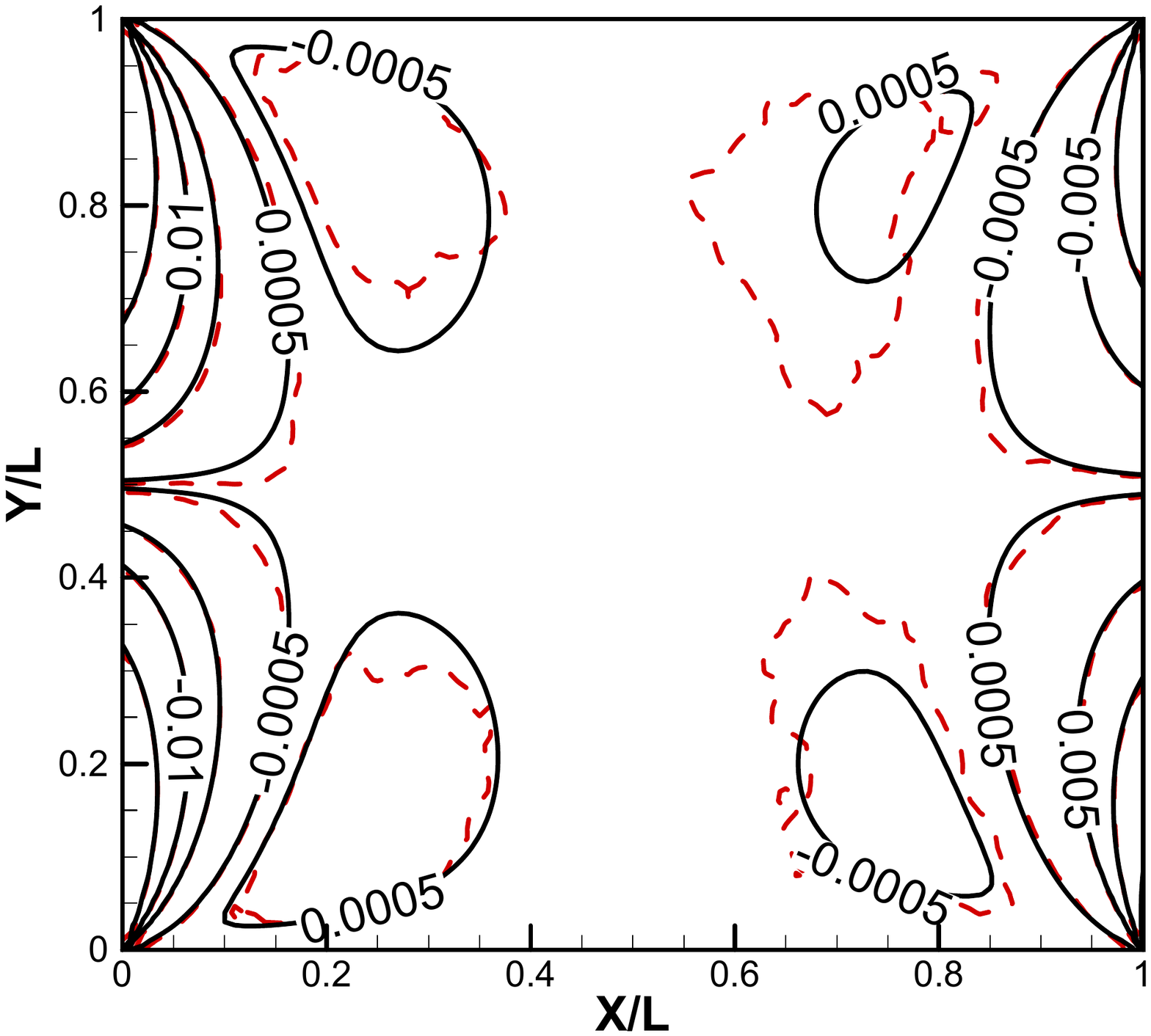}
  \end{minipage}
\caption{Normalized $U$ and $V$ velocity components, $U(X,Y)/\sqrt{2RT_L}$, $V(X,Y)/\sqrt{2RT_L}$, with diffusely reflective surfaces. $T_L:T_R:T_T:T_B=4:2:1:3$, $n_L:n_R=1:10$, and $ L/H=2.$  Dashed lines: DSMC; solid lines: analytical.}
\label{Fig:diffuse_UV}
\end{figure}

Figure \ref{Fig:diffuse_UV} shows normalized  flowfield distributions of velocity components along the X and Y directions. There are strong gradients around the four corners and they are challenging to capture. The negative U-velocity means that the gas flows from the right to the left. Around the two exit or entrance areas, the velocity profiles are relatively uniform; and in the middle, the profiles are fully developed. The V-velocity is smaller than the U-velocity component. It is quite challenging to obtain smooth V-velocity contours by the DSMC simulations because larger sampling size is needed to suppress the relatively larger statistical scatter than that of the U-velocity. However, the excellent agreements at the two exits with mild statistical scatter strongly indicate that the analytical expressions are correct.

\begin{figure}[ht]
  \begin{minipage}[l]{0.48\textwidth}
      \centering
      \includegraphics[width=3.8in]{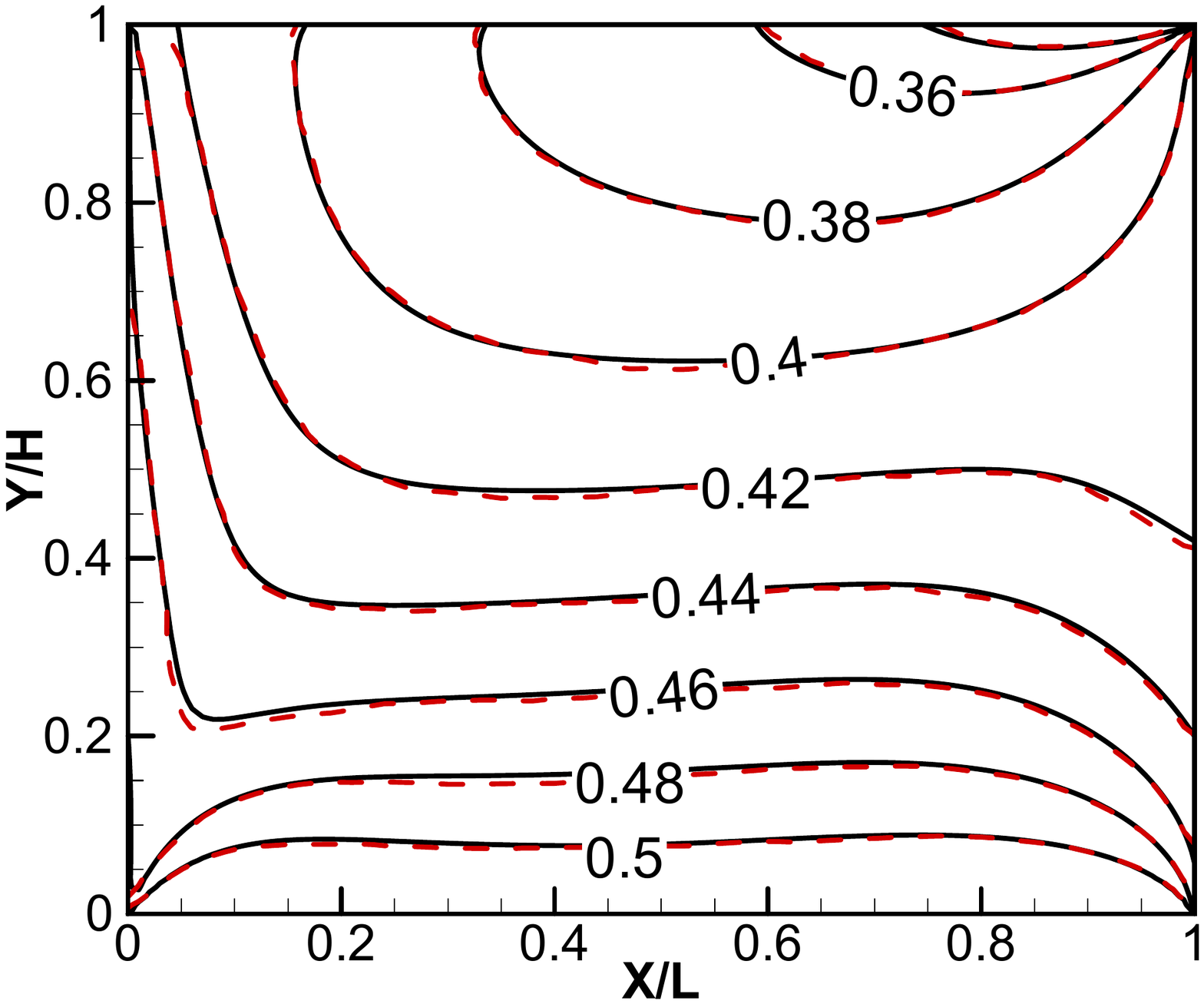}
  \end{minipage}
  \begin{minipage}[l]{0.48\textwidth}
     \centering
      \includegraphics[width=3.8in]{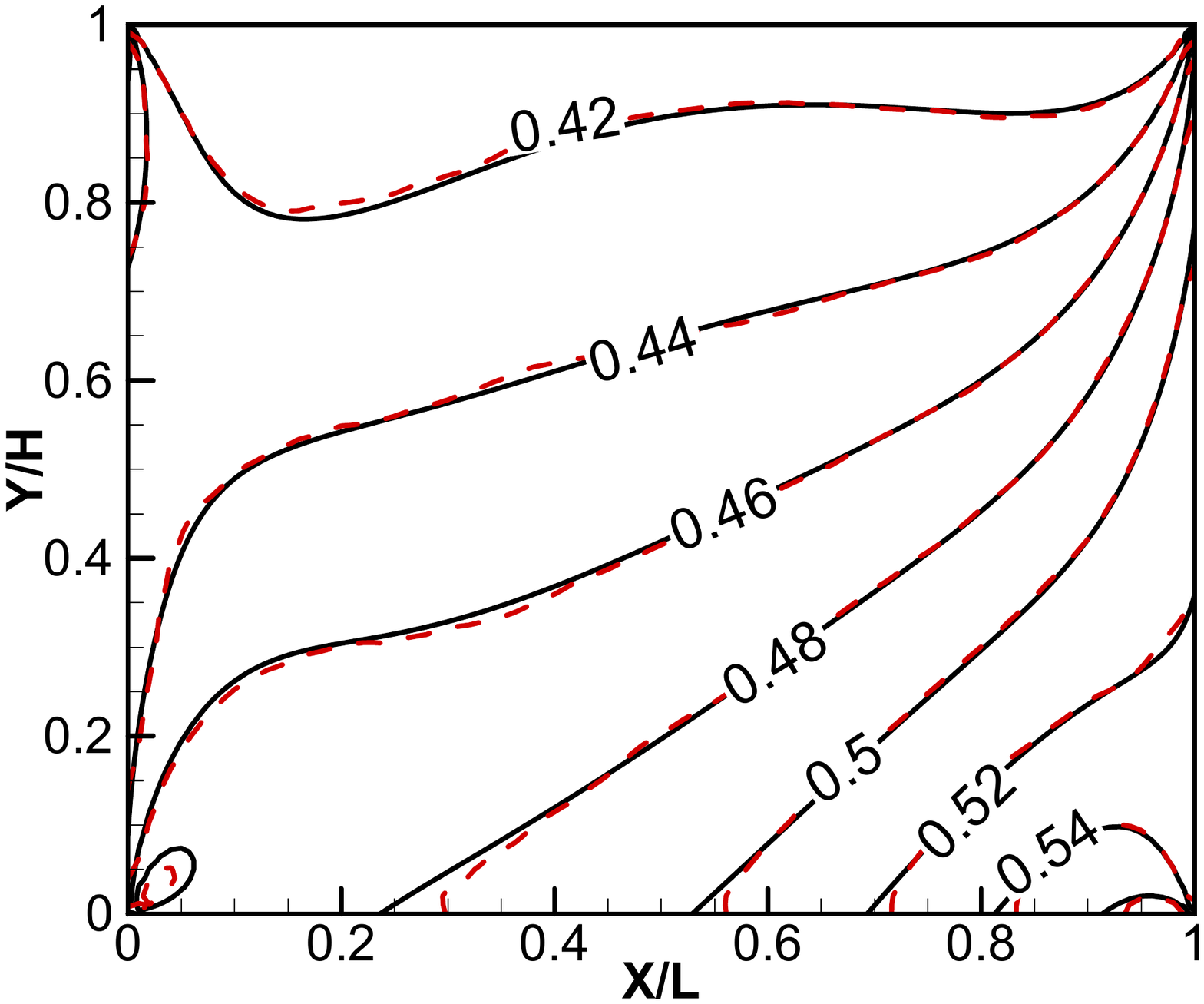}
  \end{minipage}
\caption{Translational temperatures $T_x/T_L$ and $T_y/T_L$, diffusely reflective surfaces. $T_L:T_R:T_T:T_B=4:2:1:3$, $n_L:n_R=1:10$, and $L/H=2.$  Dashed lines: DSMC; Solid lines: analytical.}
\label{Fig:diffuse_txy}
\end{figure}

Figure \ref{Fig:diffuse_txy} shows normalized translational temperature contours, $T_x/T_L$, and $T_y/T_L$, where non-equilibrium effects are appreciable, especially there are abrupt changes around the four corner regions. In the middle of the channel, the variations are mild. 

\begin{figure}[ht]
   \begin{minipage}[l]{0.48\textwidth}
      \centering
      \includegraphics[width=4.0in]{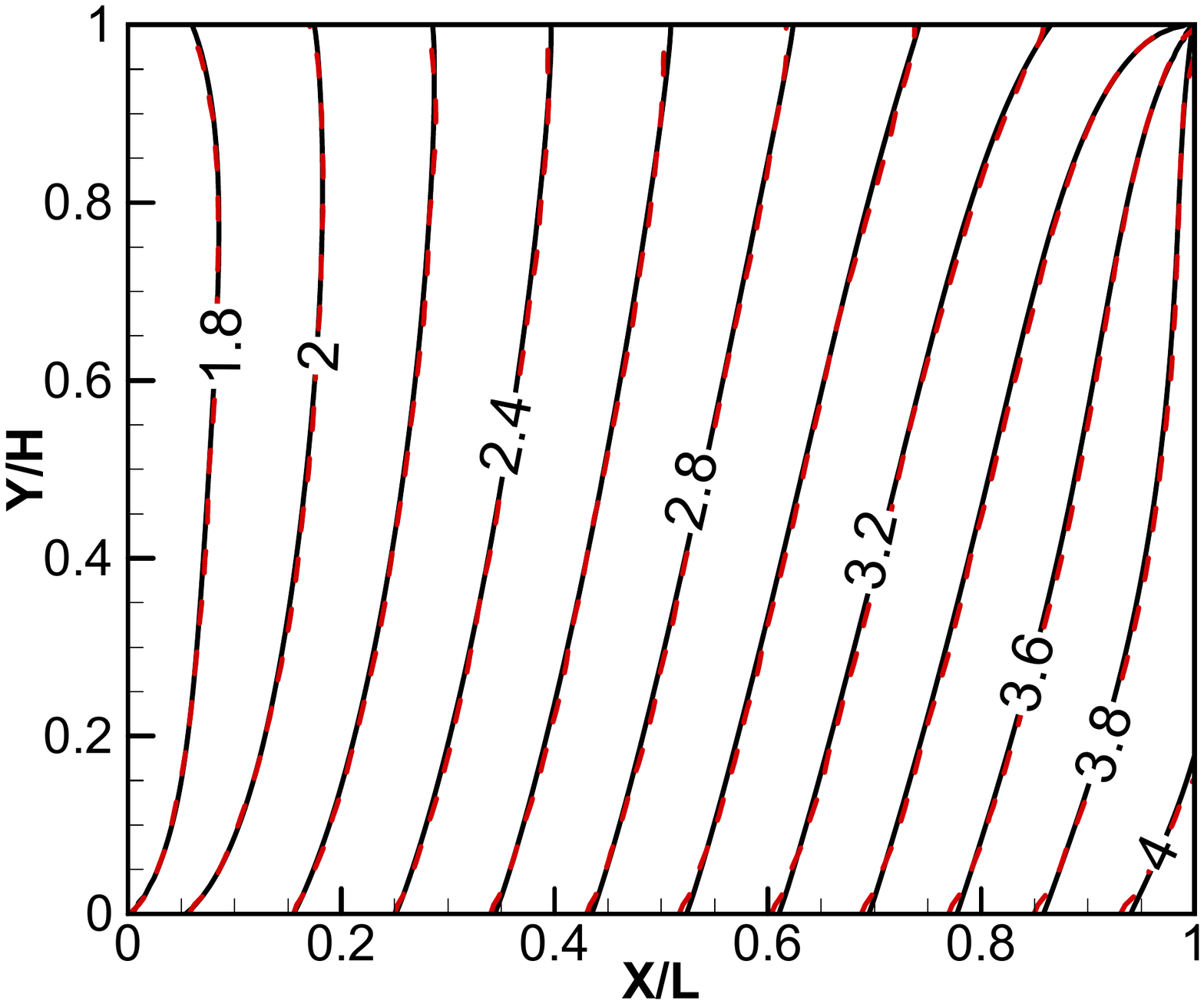}
  \end{minipage}
  \begin{minipage}[l]{0.48\textwidth}
     \centering
      \includegraphics[width=3.5in]{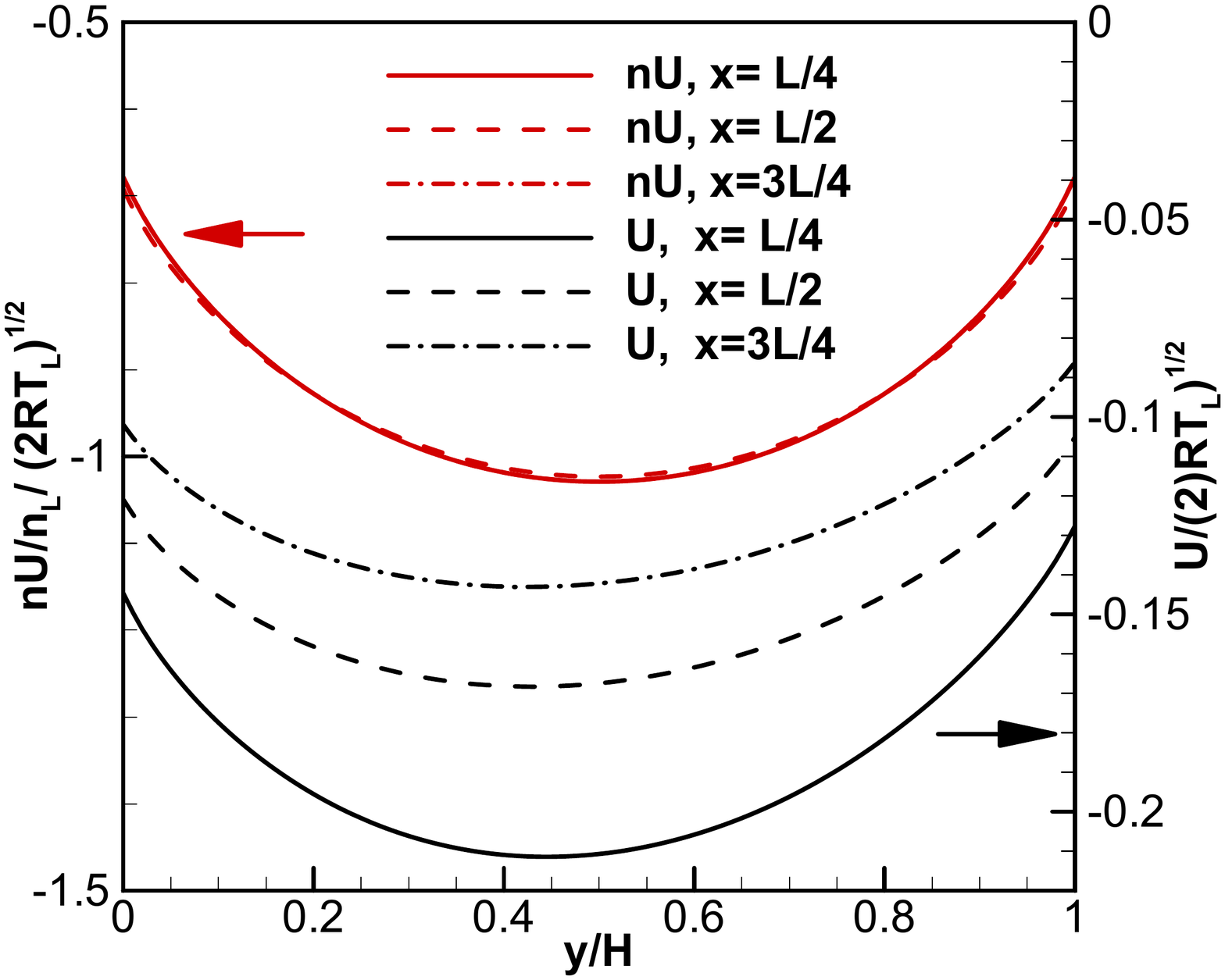}
  \end{minipage}
\caption{Left: normalized pressure field $p(X,Y)/(n_L kT_L)$, with diffusely reflective surfaces, dashed: DSMC; solid: analytical. Right: velocity and mass flux profiles at stations $x/L=0.25$, $0.50$, and $0.75$.  $T_L:T_R:T_T:T_B=4:2:1:3$, $n_L:n_R=1:10$, and $L/H=2.$}
\label{Fig:diffusep_flux}
\end{figure}

The left side of Fig. \ref{Fig:diffusep_flux} shows the pressure contours. Because $T_T$ and $T_B$ are different, those contour lines are not symmetric with respect to the channel centerline. The right side of Fig. \ref{Fig:diffusep_flux} shows the velocity and mass flux profiles at three stations $x/L =0.25$, $0.5$, and $0.75$. As shown, the flow accelerates towards the left exit, and the velocity profiles look like parabolic, very similar to a typical pressure driven Poiseuille flow. Close to the plate surface, the velocity slips are evident. At these three stations, the mass flux profiles $n(X,Y)U(X,Y)$ remain unchanged due to the mass conservation relation. 

\begin{figure}[ht]
   \begin{minipage}[l]{0.48\textwidth}
      \centering
      \includegraphics[width=3.8in]{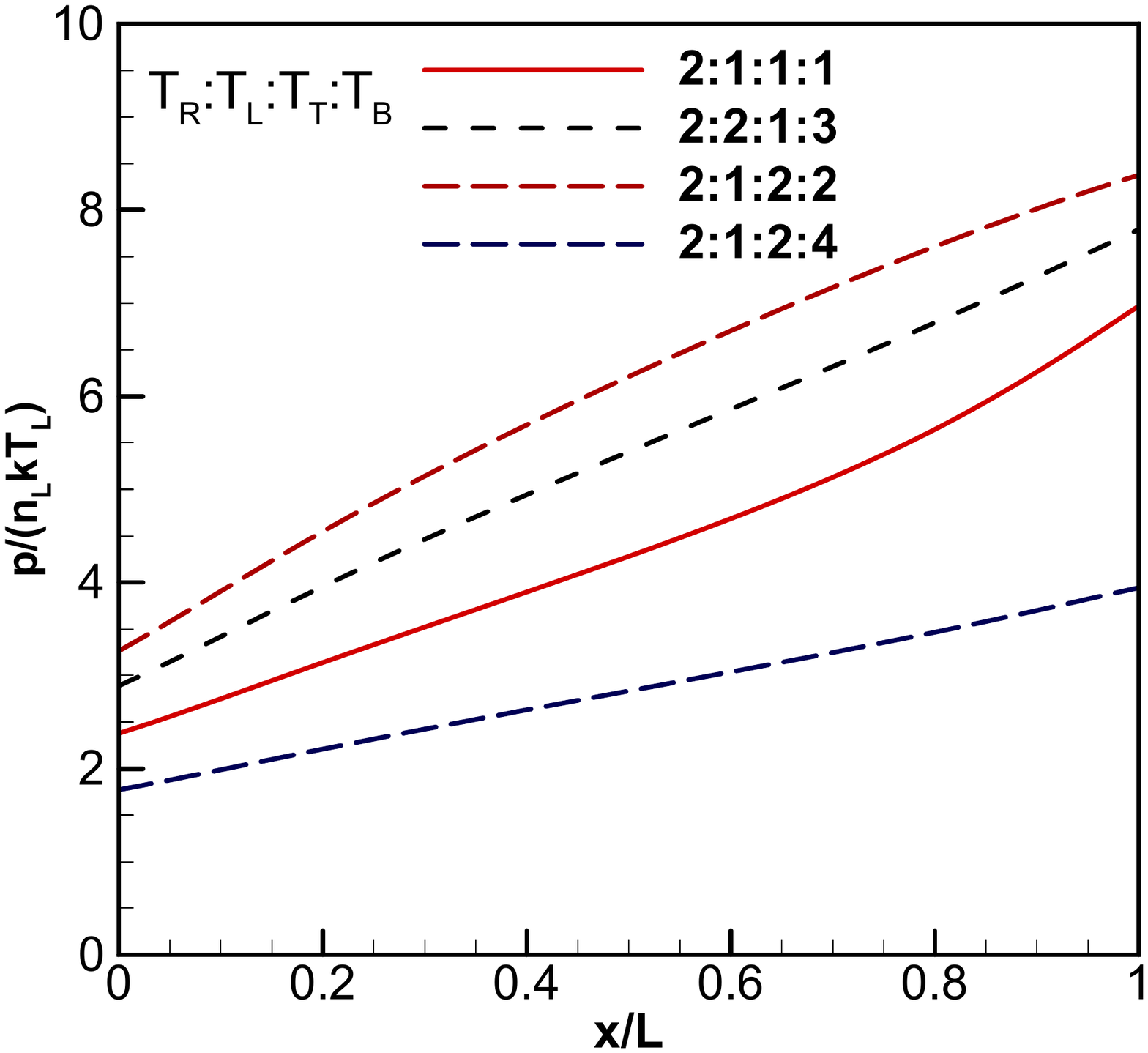}
  \end{minipage}
  \begin{minipage}[l]{0.48\textwidth}
     \centering
      \includegraphics[width=3.8in]{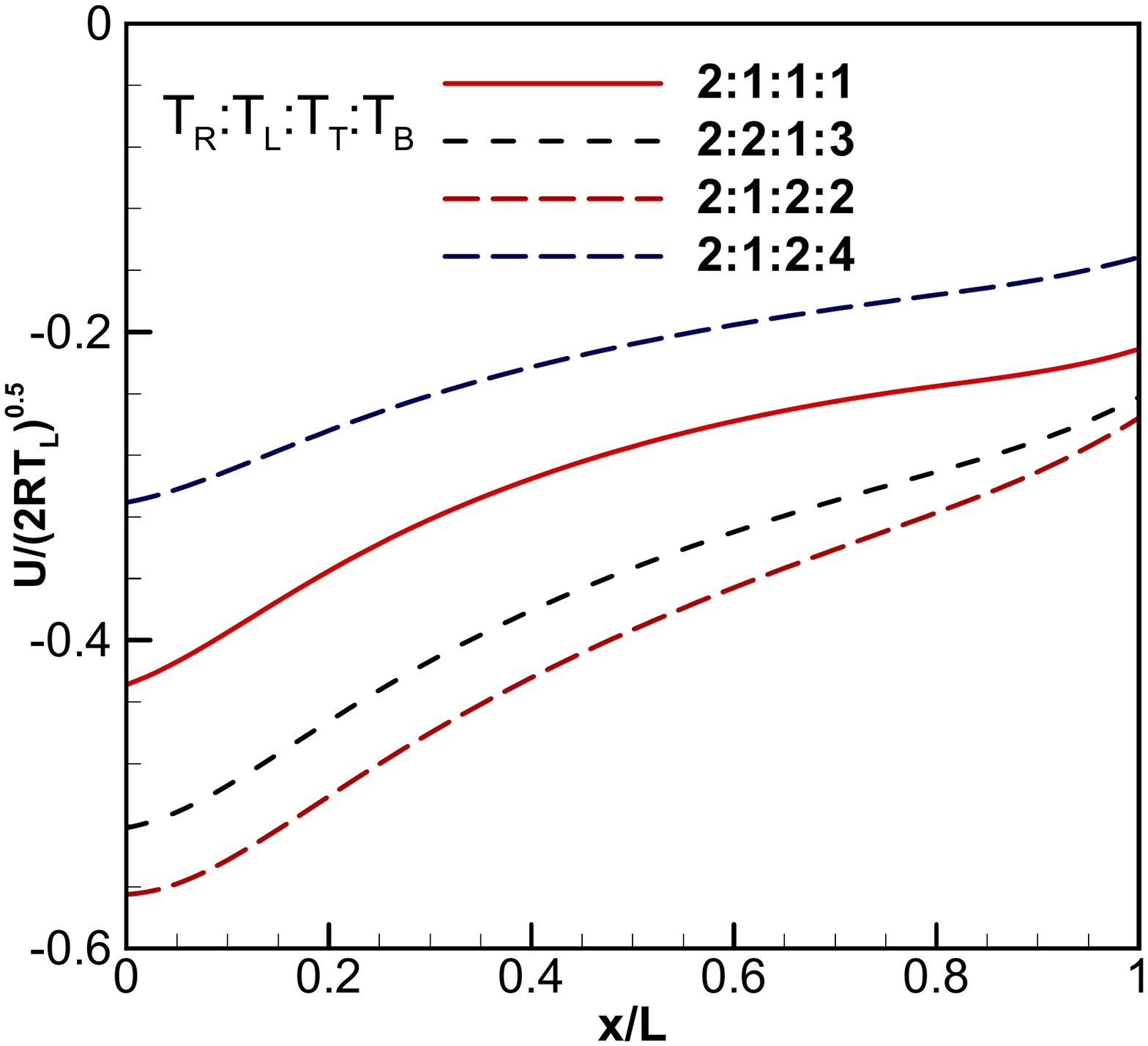}
  \end{minipage}
\caption{ Analytical solutions with diffusely reflective surfaces for the centerline pressure $p(x/L)/(n_L kT_L)$ (Left) and centerline velocity $U(x/L)/\sqrt{2RT_L}$ (Right). $n_R/n_L=5$ and $L/H=2.$ $T_L:T_R:T_T:T_B=2:1:1:1$; $2:2:1:3$; $2:1:2:2$; and $2:1:2:4$, respectively. }
\label{Fig:center_pu}
\end{figure}

It is interesting to examine how the centerline pressure  and velocity profiles change with the temperature ratios. Figure \ref{Fig:center_pu} presents those profiles of analytical solutions, where the density ratio is set as $n_R/n_L=5$ and the aspect ratio is $L/H=2$, and four different temperature $T_R:T_L:T_T:T_B$ are included. The left sub-figure is for normalized centerline pressure, and the right side is for the normalized centerline velocity. They indicate that in general the profiles are nonlinear. Also, due to different temperature ratios, either due to hot gas or hot plates, can significantly affect the profiles, e.g., their slopes and ending values. Those facts are heuristic and reflect the significance of this study.

One of our primary goals is to demonstrate that the mass flow rate is not only related to the density and temperature ratios between the two entrances or exits, but also the surface temperatures.  As well known, because the mass flux profile $n(X,Y)U(X,Y)$ is parabolic, the average value of the profile and the maximum value at the channel centerline are closely related. These two facts inspire us to study the mass flow rate by using $n(L/2,H/2)U(L/2,H/2)$, i.e., at the channel center point. 

As shown earlier in Eqn. \ref{Eqn:Uxy},  the surface virtual density is crucial to compute the local mass flux, $n(X,Y)U(X,Y)$, and we need to approximate it in the first step. Inspired by the right sub-figure in Fig. \ref{Fig:nt_nb}, if $T_T=T_B =(T_L+T_R)/2$, the top and bottom plates shall share an identical virtual density profile which is almost linear. This virtual density profile is renamed as $n_w(x)$, and it is approximated with the following formula:
\begin{equation}
     n_w(x)= \frac{n_L+n_R}{2} (a+bx); \mbox{or}, n_w(\alpha) =  \frac{n_L+n_R}{2} \left( a + b \left[ \frac{L}{2H} - \frac{ \cos\alpha}{2\sin \alpha} \right] \right),  \alpha_L < \alpha < \pi - \alpha_L,
\label{eqn:nwtb}
\end{equation}
where $a$ and $b$ are two determinable coefficients, e.g., by using the right sub-figure in Fig. \ref{Fig:nt_nb}.

By the assumption shown as Eqn. \ref{eqn:nwtb}, and the following integral,
\begin{equation}
     \int \frac{ \cos^2 \alpha }{\sin \alpha} d\alpha =   \cos \alpha  - \frac{1}{2} \mbox{ln} \frac{1+ \cos \alpha}{1- \cos \alpha }+C,
\end{equation}
it can be shown that the local mass flux at the center point $(L/2, H/2)$ has the following expression:
\begin{equation}
      (nU)(\frac{L}{2},\frac{H}{2})= \frac{(n_L+n_R) \sqrt{2RT_w}}{8 \sqrt{\pi}} \left[    \cos \alpha_0 +  \frac{1}{2} \mbox{ln} \frac{ 1- \cos\alpha_0 }{1+ \cos \alpha_0} \right] - \frac{ \sqrt{2RT_R} }{2 \sqrt{\pi}} n_R \sin \alpha_0  + \frac{ \sqrt{2RT_L} }{2 \sqrt{\pi}} n_L \sin \alpha_0,
\end{equation}
where $\alpha_0 = \mbox{atan} (H/L)$, the top and bottom plates have the same temperature $T_w=(T_L+T_R)/2$. Hence, the normalized mass flux is: 
\begin{equation}
     \frac{Q}{n_L H \sqrt{2RT_L}} \sim \frac{b}{4}  \left(1+ \frac{n_R}{n_L} \right)\sqrt{ \frac{T_w}{T_L}} \left(   \cos \alpha_0 + \frac{1}{2} \mbox{ln} \frac{ 1 - \cos \alpha_0}{1+ \cos\alpha_0} \right)  -  \sqrt{\frac{T_R}{T_L}} \frac{n_R}{n_L} \sin \alpha_0 +  \sin \alpha_0.
\label{eqn:massflows}
\end{equation}
Equation \ref{eqn:massflows} includes four non-dimensional parameters, $L/H$, $n_R/n_L$, $T_w/T_L$, and $T_R/T_L$. This expression shows the pressure ratio $P_L/P_R = (n_L T_L)/(n_R T_R)$ is not a non-dimensional factor for the mass flow rate through a micro-channel. However, Berman's formulas \cite{Berman, Bermanerror}  for collisionless gas flow through a planar micro-channel uses the pressure difference  (or the density difference)  as a factor to determine the mass flow rate. 

For a long micro-channel, e.g, $H/L <0.1$, the above equation can further degenerate to: 
\begin{equation}
     \frac{Q}{n_L H \sqrt{2RT_L}} \sim \frac{b}{4} \left(1+ \frac{n_R}{n_L} \right)\sqrt{ \frac{T_w}{T_L}} \mbox{ln} \frac{H}{L}  -  \sqrt{\frac{T_R}{T_L}} \frac{n_R}{n_L} \frac{H}{L} +  \frac{H}{L}.
\label{eqn:massflows2}
\end{equation}
When the left and right exits have identical number densities and temperatures, we have $b=0$ and then Eqns. \ref{eqn:massflows} and \ref{eqn:massflows2} correctly predict a zero mass flow rate. 

\begin{figure}[ht]
   \begin{minipage}[l]{0.48\textwidth}
      \centering
      \includegraphics[width=3.8in]{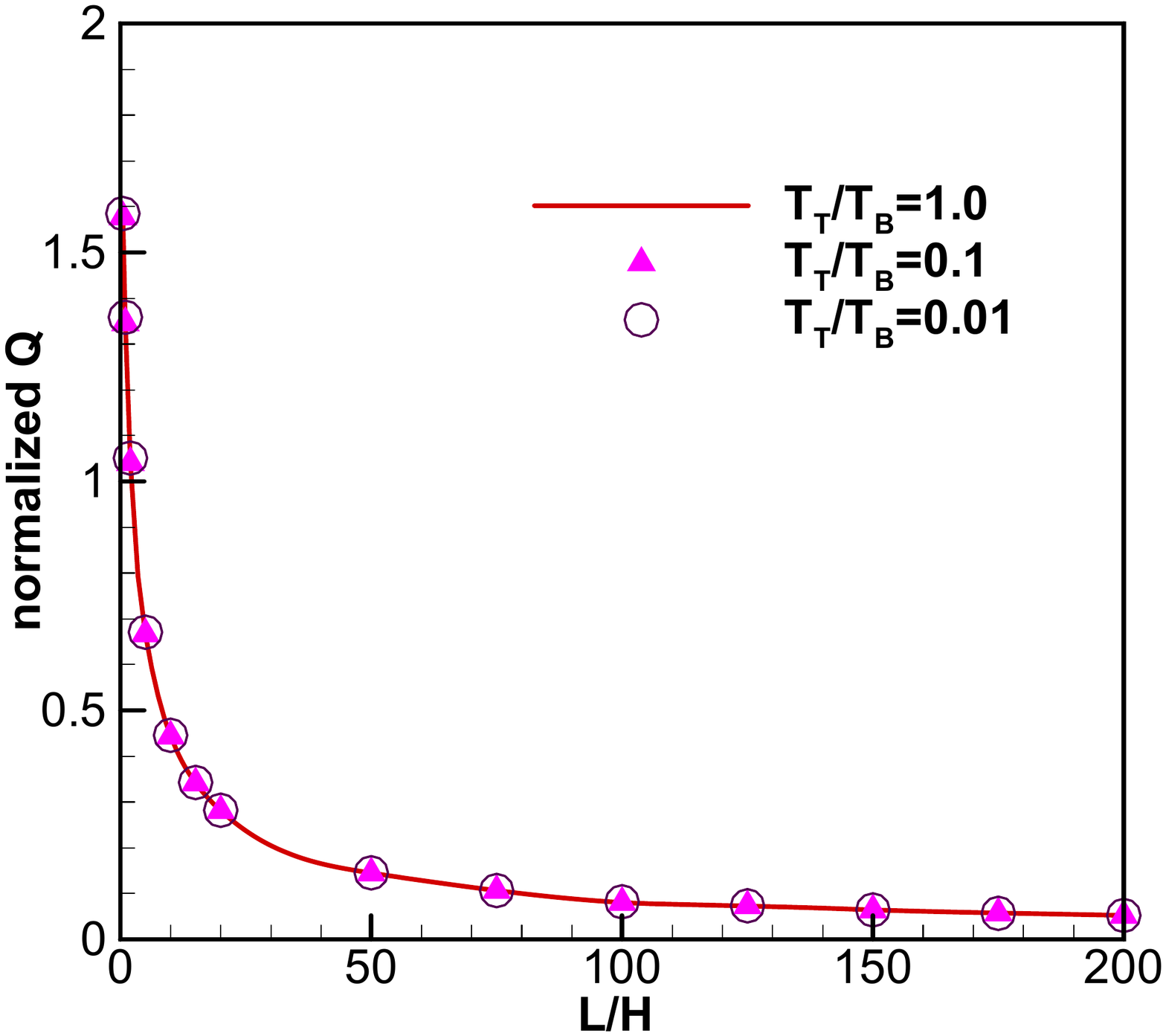}
  \end{minipage}
  \begin{minipage}[l]{0.48\textwidth}
     \centering
      \includegraphics[width=3.8in]{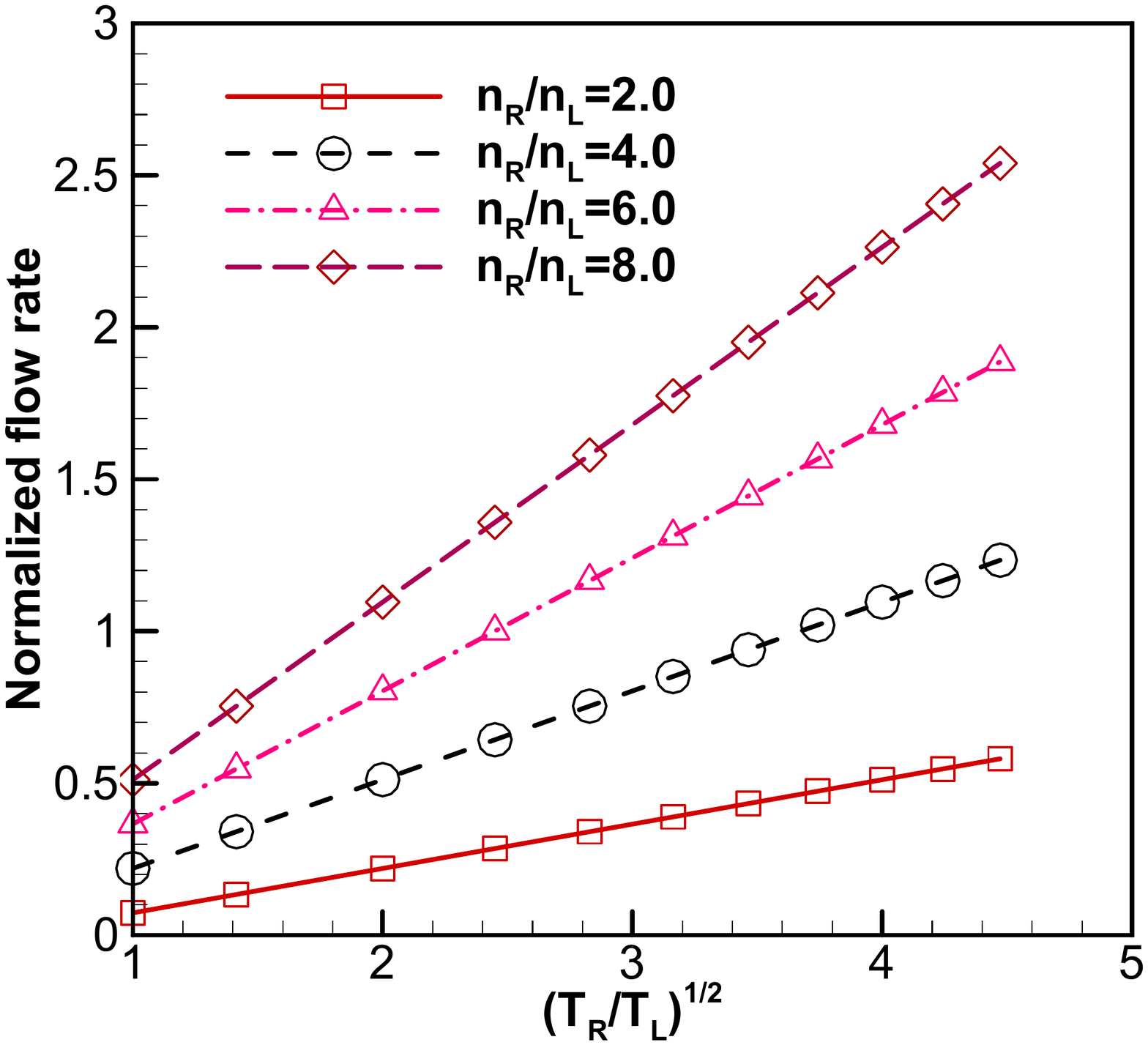}
  \end{minipage}
\caption{Left: $L/H$ ratio effect on mass flow rate through a channel with diffusely reflective surfaces, $T_L:T_R=2:1$ and $n_L:n_R=1:10$. Right: temperature ratio effect on mass flow rate, $L/H=10$, and $T_T=T_B = (T_L+T_R)/2.$  The mass flow rate is computed at $x/L=0.5$  for both cases}.
\label{Fig:nu_TL_TR}
\end{figure}

Figure \ref{Fig:nu_TL_TR} shows the normalized flow rate profiles, $Q/(n_L \sqrt{2RT_L} H)$, with different factors. The left side illustrates that the flow rate decreases with increasing $L/H$ because a relatively longer channel slows down the molecules more effectively. In this left sub-figure, three different plate temperature ratios are used, and, as shown for long micro-channels, the temperature ratio effects are not very appreciable and the difference is less than 0.1\%.
The  right side of this figure shows the normalized mass flow rate, $Q/(n_L \sqrt{2RT_L} H)$, changing with different temperature ratios, and four different scenarios with different density ratios are displayed. As illustrated by Eqn. \ref{eqn:massflows}, with  a constant surface temperature $T_w = (T_L+T_R)/2$, if the linear virtual surface density profile assumption is reliable (i.e, $b$ is a constant) , the normalized mass flow rate shall have a simple linear relation with  $\sqrt{T_w/T_L}$. The four straight lines in the right sub-figure clearly confirm this relation.

\section{CHANNELS WITH SPECULARLY REFLECTIVE SURFACES}
\label{sec:specular}
We can also  study free molecular flows through a channel with specularly reflective surfaces by constructing the VDFs and velocity phases for the two plate surfaces. From the gaskinetic theory,  the surface temperatures $T_T$ and $T_B$ do not have any contribution to the surface and flowfield properties. The shear stress and heat flux at the surface are zero. The VDFs at the left and right exits still can be well approximated as Maxwellian for free molecular flows.

\begin{figure}[ht]
\centering
\includegraphics[trim=0 90 0 80, clip, width=5in, height=2.3in]{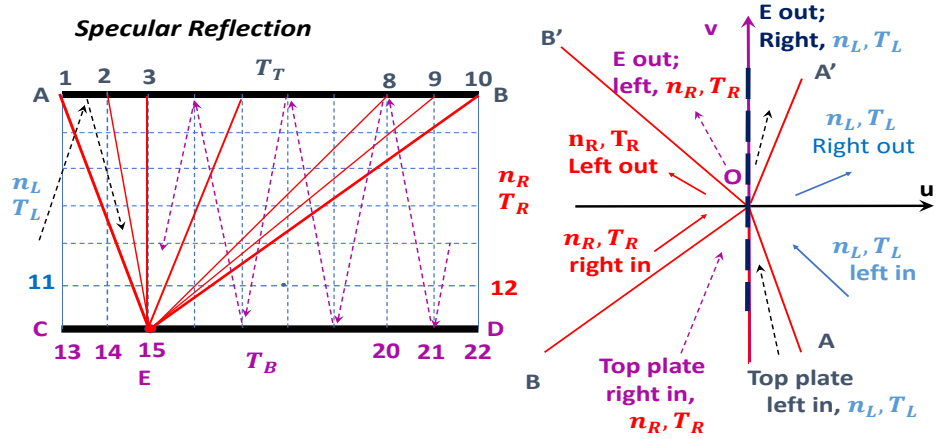}
\caption{Left: a sketch for specularly reflective channel surface property computations; Right: velocity phases for surface Point $E$.}
\label{fig:illustration_specular_surf}
\end{figure}

Figure \ref{fig:illustration_specular_surf} can aid studying the velocity phase for Point $E$ at the bottom plate. There are particles from multiple sources moving toward Point $E$. The first group of molecules enter the channel from the left entrance. They move towards Point $E$ and the related velocity phase is inside $\angle AOu$ in Quadrant IV. These molecules bounce off from Point $E$ and leave the channel through the right exit with the related velocity located inside $\angle A'Ou$ (which is symmetric to $\angle AOu$) in Quadrant I. Molecules entering the channel from the right entrance and moving towards Point $E$ are represented by angle $\angle BO(-u)$ in Quadrant III, and they bounce off from $E$ and move to the left exit, and the related velocity phase shall be $\angle B'O(-u)$ (symmetric to $\angle BO(-u)$) in Quadrant II.  

An equivalent approach to understand this velocity phase construction is to replace those particles bouncing off Point $E$ with two imaginary groups, which are equivalently coming under the plate $CD$ at velocity ``mirrored'' from the actual molecules coming from the left and right entrances. By doing this, the plate $CD$ can be neglected, and there are four groups of molecules coming to Point $E$: within $\angle AOu$ and within $\angle A'Ou$ (mirrored group); within  $\angle BO(-u)$ and within $\angle B'O(-u)$ (mirrored group).    

There are also two groups of molecules moving toward Point $E$ from the top plate. They are reflected molecules, either from plate segment 3-B or A-3. Specular reflections only reverse the velocity component along the normal direction but the tangent velocity component is unchanged. It is understandable that there are many molecule ``trains'' whose velocities are actually Maxwellian, and form a ray in the velocity phase. This ray can trace back to the left or the right entrances. In Fig. \ref{fig:illustration_specular_surf}, molecules moving from surface segment 3-A towards $E$ can only come from the left entrance, and they are represented inside $\angle AO(-v)$ in Quadrant IV, and they bounce off from $E$ and form an area $\angle A'Ov$ (symmetric to $\angle AO(-v)$) in Quadrant I. Similarly, molecules moving from the segment 3-B towards $E$ form $\angle BO(-v)$ in Quadrant III, and they leave Point $E$ and move to the left exit, forming an area $\angle B'Ov$ (symmetric to $\angle BO(-v)$) in the Quadrant II.

The above classification yields simple conclusions about the velocity phase for Point $E$, as shown on the right side of Fig. \ref{fig:illustration_specular_surf}. In  Quadrants I and IV with positive $u$, the VDFs are Maxwellian and characterized by $n_L$ and $T_L$; while the VDFs in Quadrants II and III with negative $u$ are also Maxwellian but characterized by $n_R$ and $T_R$. These four quadrants are divided by two rays: $\theta = 90^\circ$, and $\theta = 270^\circ$, which are missing because no molecule from the two entrances can bounce into and out of Point $E$, vertically along Segment E-3. Molecules with such velocity directions can never enter the channel from the two entrances. However, these two missing rays in the velocity phase do not affect the integration to obtain the surface properties according to the standard gaskinetic theory, because the integral area of their occupied solid angles is zero.  Following the same procedures, the VDF and velocity phase for any point on the top and bottom surfaces shall be the identical.

\begin{figure}[ht]
\centering
\includegraphics[trim=0 90 0 80, clip, width=5in, height=2.3in]{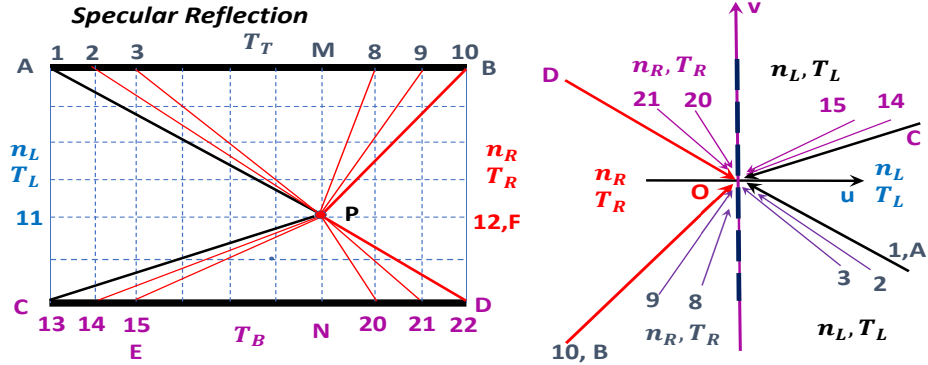}
\caption{Left: a sketch for specularly reflective channel flowfield property computations; Right: velocity phases for flowfield Point $P(X,Y)$.}
\label{fig:illustration_specular_field}
\end{figure}

After determining the VDFs and velocity phases for all surface points, we can proceed to construct the VDF and velocity phase for a flowfield point $P(X,Y)$ in Fig. \ref{fig:illustration_specular_field}, and compute the flowfield properties at that point. The velocity phase for this point is shown on the right side of this figure. Region $\angle AOC$ in the velocity phase represents molecules entering the channel and directly moving towards point $P(X,Y)$ from the left entrance; Region $\angle DOB$ in the velocity phase represents molecules entering the channel and directly moving towards point $P(X,Y)$ from the right entrance. Neither $\angle AOC$ nor $\angle DOB$ needs to be symmetric about the u-axis. Region $\angle COv$ in the velocity phase is formed by those molecules bouncing off plate segment $CN$; and region $\angle AO(-v)$ represents molecules bouncing off plate segment AM. Regions $\angle AO(-v)$, $\angle COV$ and $\angle COA$ form Quadrants I and IV, and the corresponding VDF is Maxwellian and characterized by $n_L$ and $T_L$. With the same approach, we can conclude that the VDF inside Quadrants II and III is also Maxwellian but characterized by $n_R$ and $T_R$. The $90^\circ$ and $270^\circ$ rays in the velocity phase are still missing, however, neither of them has any effect on the integration for the flowfield properties at point $P(X,Y)$. Another conclusion that can be drawn is that the velocity phase for any flowfield point shall be identical to that of point $P$, regardless of the channel aspect ratio.  

In conclusion, the VDFs and velocity phases for all the flowfield and surface points are identical: one Maxwellian VDF (characterized by $n_L$ and $T_L$) for Quadrants I and IV, and another Maxwellian VDF (characterized by $n_R$ and $T_R$) for Quadrants II and IV. As a consequence, the flowfield properties must be identical and uniform everywhere. The geometry ratio $L/H$ and the surface temperature ratios $T_T/T_L$ and $T_B/T_L$ do not appear in the VDFs and velocity phases. No matter how long the channel is, the mass flow rate through the channel is fixed, due to the zero surface friction force. By taking the whole channel as a control volume, molecules entering from one entrance must leave from the other exit, and they shall never bounce back and leave from the same entrance.   

Having the local VDFs and velocity phases obtained, we can integrate VDFs in the velocity phases for the local surface and flowfield properties by using the gaskinetic theory. For convenience, two non-dimensional parameters are defined, $\epsilon_n \equiv n_R/n_L$, and $\epsilon_T \equiv T_R/T_L$. With them, the surface pressure coefficient  can be expressed as:
\begin{equation}
    C_{p,s}(X) = \frac{p_w}{n_L kT_L}  =  \frac{1}{2} +  \frac{1}{2} \epsilon_n \epsilon_T.
\end{equation}
The flowfield properties are: 
\begin{equation}
     \frac{n(X,Y)}{n_L}  = \frac{1}{2} +   \frac{\epsilon_n}{2}; \frac{T(X,Y)}{T_L} = - \frac{U^2}{ 3RT_L}  +  \frac{T_y}{T_L}; 
\label{eqn:specular_nT}
\end{equation}     
\begin{equation}
    \frac{U(X,Y)}{\sqrt{2RT_L}} =   \frac{ 1- \epsilon_n  \sqrt{ \epsilon_T  }  }{ \sqrt{\pi} (1+ \epsilon_n) }; V(X,Y)=0;
\label{eqn:specular_uv}
\end{equation}
\begin{equation}
       \frac{T_x(X,Y)}{T_L} = - \frac{U^2}{ RT_L}  + \frac{T_y}{T_L}; 
       \frac{T_y(X,Y)}{T_L} =\frac{T_z(X,Y)}{T_L} =  \frac{ 1 + \epsilon_n \epsilon_T }{2}  \frac{n_L}{n}.
\label{eqn:spec_TxTy}
\end{equation}
Equations \ref{eqn:specular_nT} and \ref{eqn:spec_TxTy} indicate that there are strong thermal non-equilibrium effect on temperatures, and such non-equilibrium is not related to the aspect ratio of the channel. $T_y$ and $T_z$ are equal and always the largest, then is the averaged temperature $T$, and the translational temperature along the X-direction $T_x$ is always the smallest. 


Compared with the diffusely reflective plate results having four non-dimensional parameters, the above results are simpler. The distributions have uniform values everywhere.

\begin{figure}[ht]
  \begin{minipage}[l]{0.48\textwidth}
      \centering
      \includegraphics[width=3.8in]{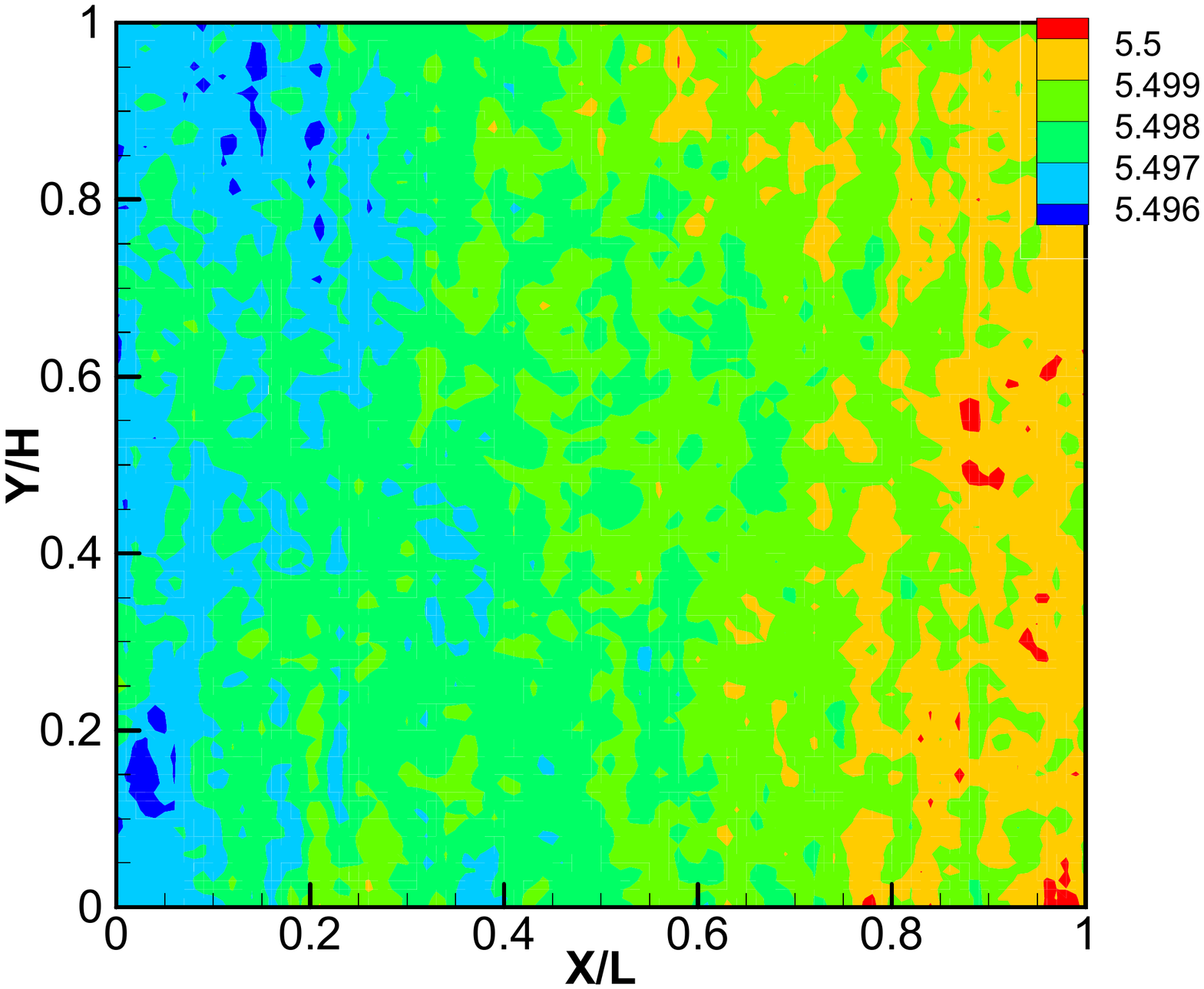}
  \end{minipage}
  \begin{minipage}[l]{0.48\textwidth}
     \centering
      \includegraphics[width=3.8in]{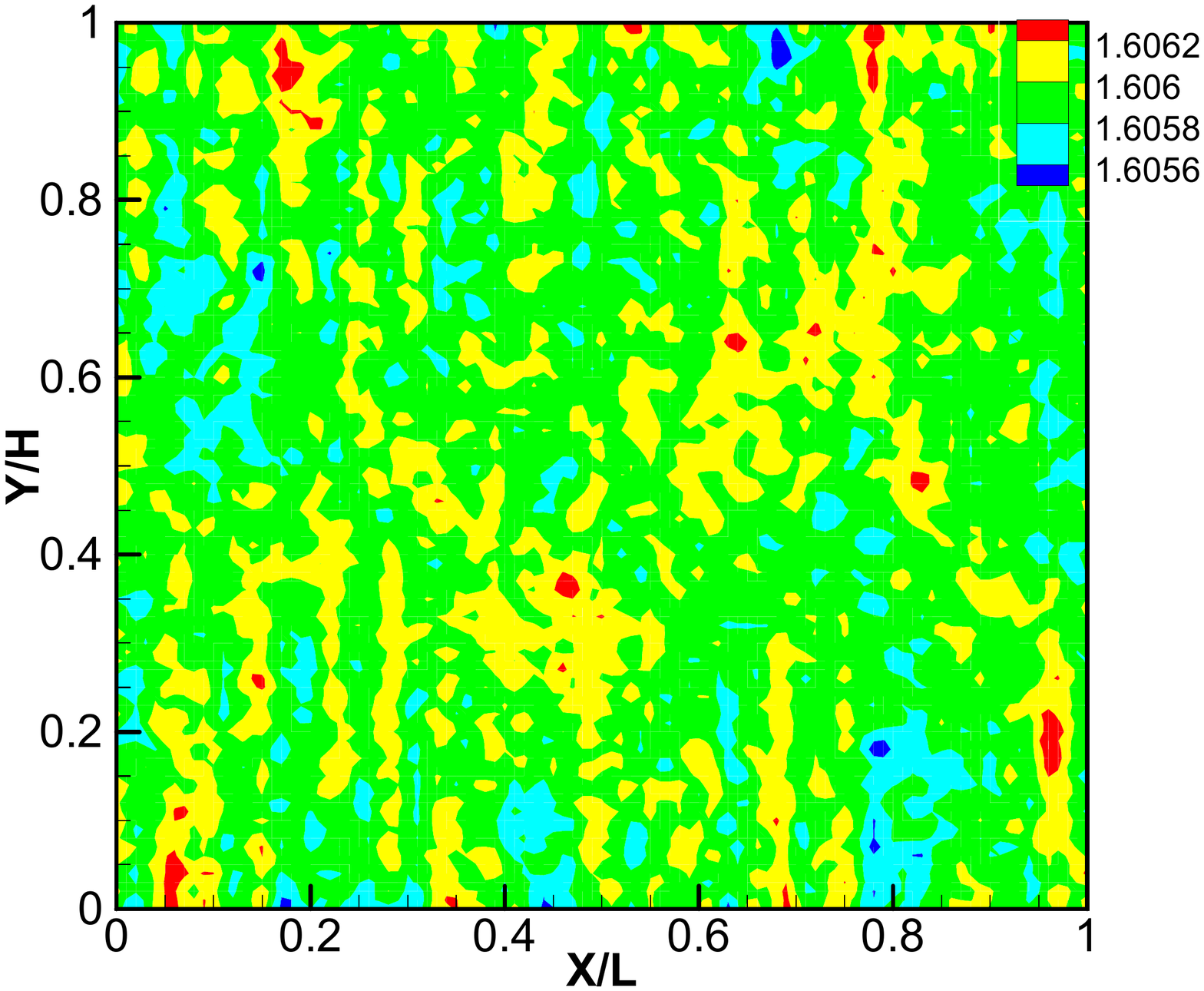}
  \end{minipage}  
\caption{DSMC simulation results of normalized density and translational temperature, specularly reflective channel surfaces. $T_L:T_R=1:2$, $n_L:n_R=1:10$. Analytical results ( computed with Eqn. \ref{eqn:specular_nT}): $n(X,Y)/n_L =5.5$, and $T(X,Y)/T_L = 1.6062$. }
\label{Fig:nt_specular}
\end{figure}

\begin{figure}[ht]
  \begin{minipage}[l]{0.48\textwidth}
      \centering
      \includegraphics[width=3.8in]{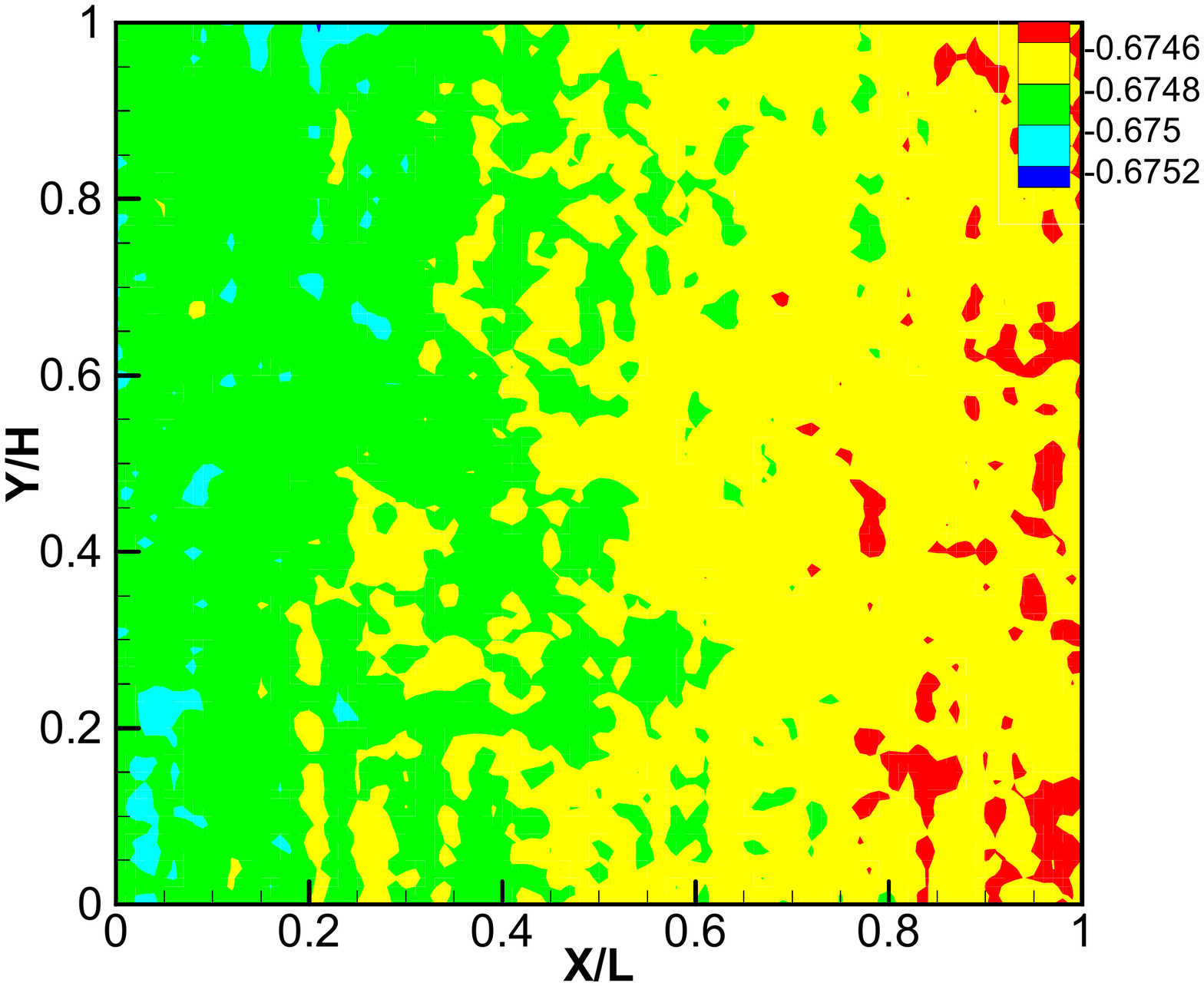}
  \end{minipage}
  \begin{minipage}[l]{0.48\textwidth}
     \centering
      \includegraphics[width=3.8in]{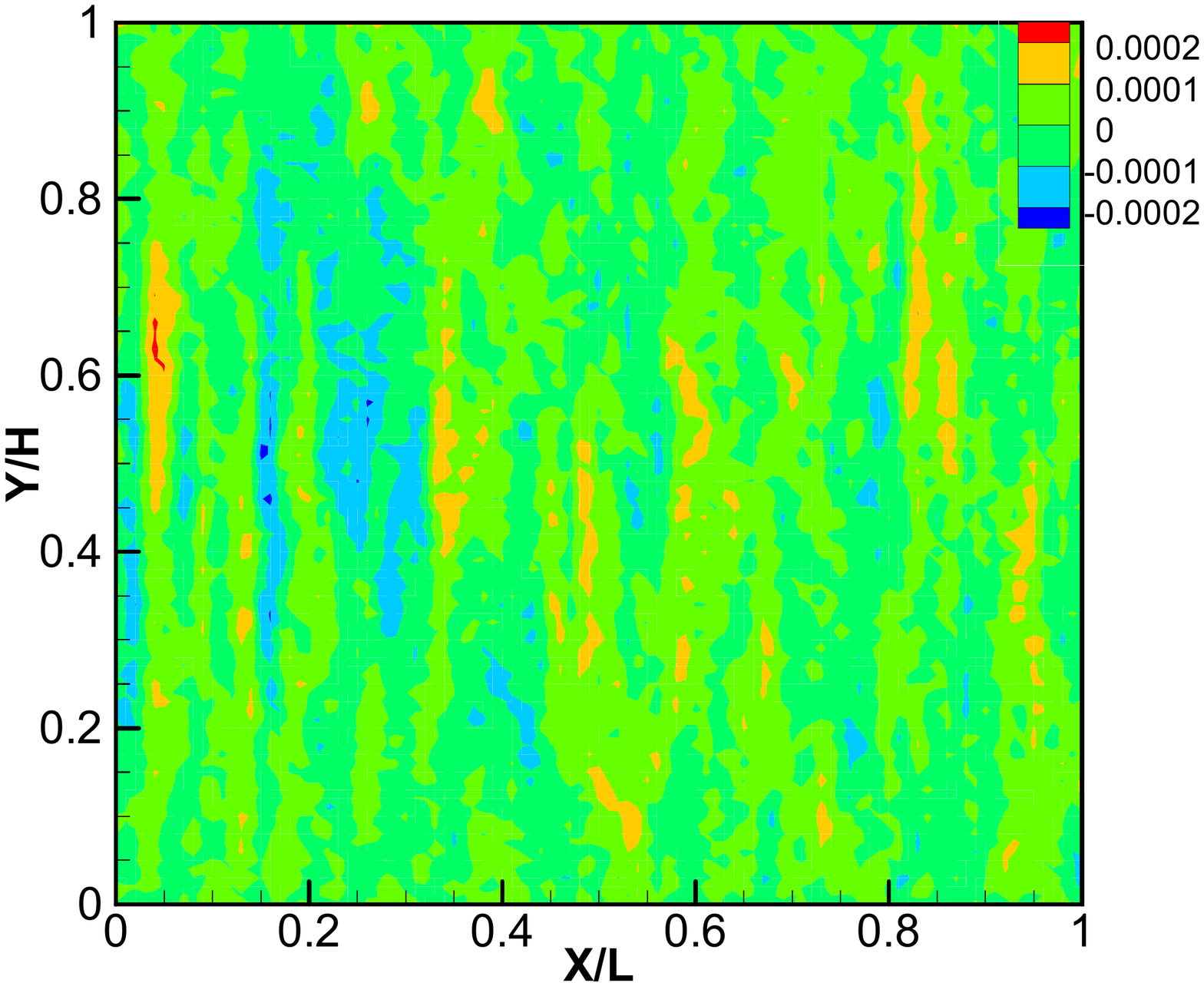}
  \end{minipage}
\caption{DSMC simulation results of  normalized velocity components, specularly reflective channel surfaces. $T_L:T_R=1:2$, $n_L:n_R=1:10$. Analytical results (computed with Eqn. \ref{eqn:specular_uv}): $U(X,Y)/\sqrt{2RT_L} = -0.675$, and $V(X,Y)/\sqrt{2RT_L} = 0.0$.}
\label{Fig:UV_specular}
\end{figure}

\begin{figure}[ht]
  \begin{minipage}[l]{0.48\textwidth}
      \centering
      \includegraphics[width=3.8in]{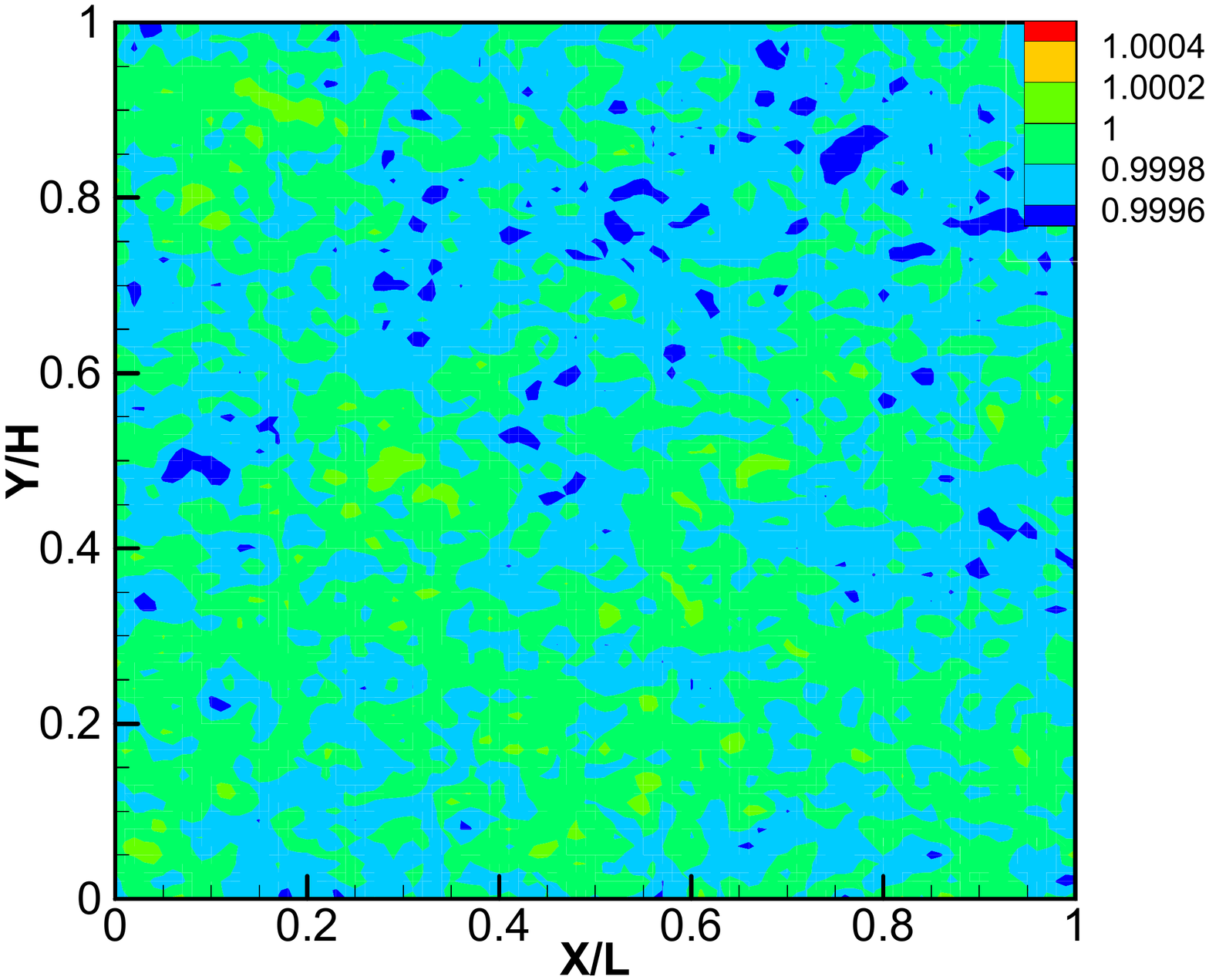}
  \end{minipage}
  \begin{minipage}[l]{0.48\textwidth}
     \centering
      \includegraphics[width=3.8in]{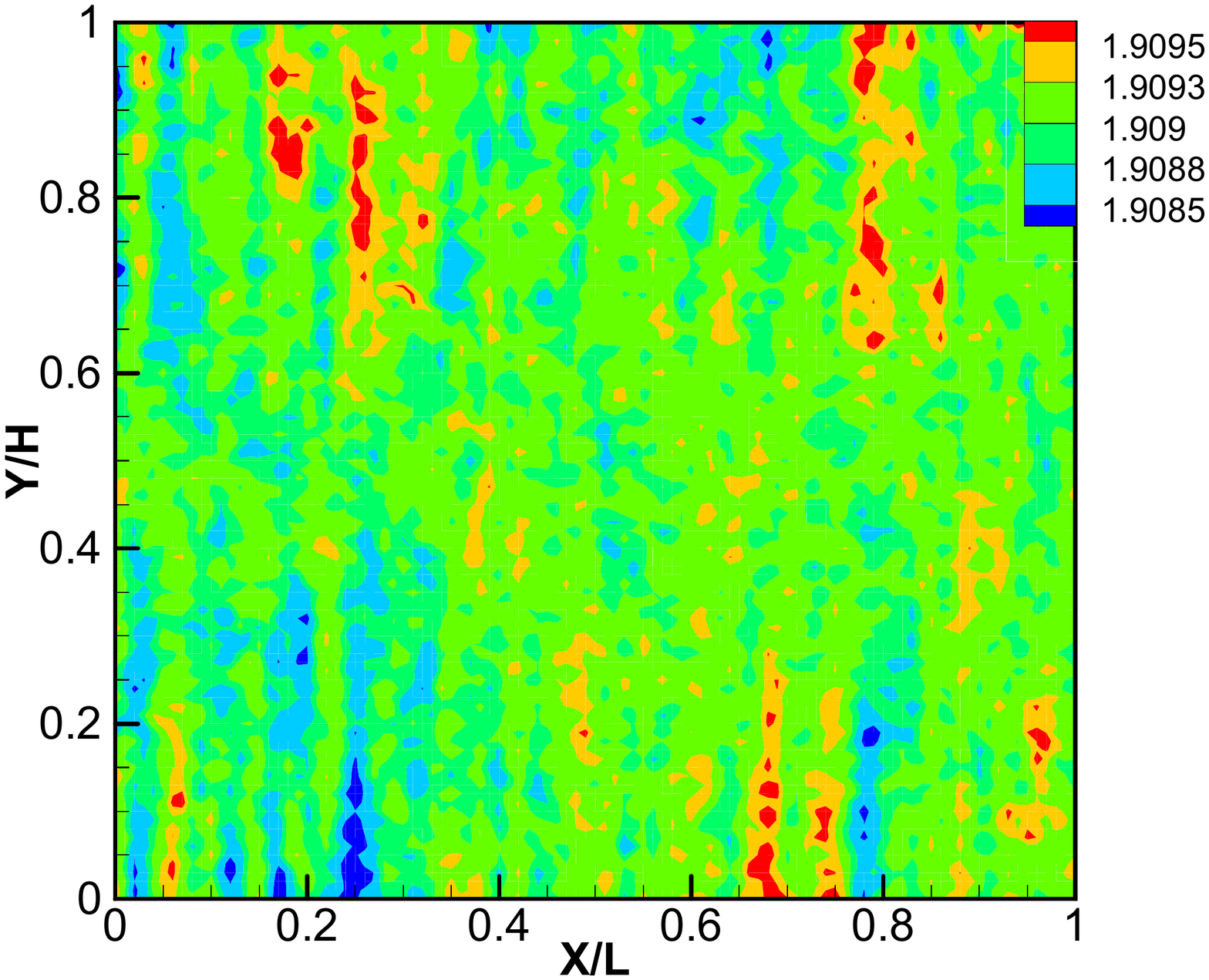}
  \end{minipage}
\caption{DSMC simulation results of normalized translational temperature components, $T_x(X,Y)/T_L$ and $T_y(X,Y)/T_L$, specularly reflective surfaces. $T_L:T_R=1:2$, $n_L:n_R=1:10$. Analytical results ( computed with Eqn. \ref{eqn:spec_TxTy}): $T_x(X,Y)/T_L = 1.00037$, and $T_y(X,Y)/T_L = 1.9091$.   }
\label{Fig:TxTy_specular}
\end{figure}

To validate the above analytical formulas for a micro-channel with specularly reflective surfaces, DSMC simulations are performed and the results are compared. Figure \ref{Fig:nt_specular} includes the simulation results of density and averaged temperature; Fig. \ref{Fig:UV_specular} for velocity components; and Fig. \ref{Fig:TxTy_specular} for translational temperature  components $T_x$ and $T_y$. The corresponding exact analytical values are also included in the captions. Even though there are still statistical scatters in the simulation results, the deviations from the analytical values are minor, in general, and the differences are on the 4th effective digit number, or about $0.01\%$. These discrepancies can further reduce if the sampling process continues. Figure \ref{Fig:TxTy_specular} also indicates that $T_x$ and $T_y$ can be quite different and the difference is about $90\%$ for this parameter combination. As shown, the formulas for the specularly reflective surface plates are also validated.

Correspondingly, the normalized mass flow rate through a planar micro-channel with specularly reflective plate surfaces is:
\begin{equation}
     \frac{Q}{n_L  H \sqrt{2RT_L}} =  \frac{ 1 - \epsilon_n \sqrt{ \epsilon_T} }{2 \sqrt{\pi}}.
\end{equation}
As shown, different from the diffusely reflective surface scenario, the normalized mass flow rate for collisionless flows through a specularly reflective micro-channel does not include the aspect ratio and surface temperatures. The above formula also indicates that for collisionless flows, even if the pressures at the two entrances are the same, i.e., $\epsilon_n \epsilon_T=1$, it is still possible to have non-zero flux through the channel. This is different from many past theoretical results, which predict that the mass flow rate shall be proportional to the pressure difference, such as Berman \cite{Berman,Bermanerror}.

{
Before we conclude this paper, few remarks are offered here. First, following the same vein, when the collisionless gases at the left and right exits have different VDFs, or the micro-channel has different but simple shapes like a long capillary, the above procedures dealing with reflective surfaces can still apply and the results shall be similar. All the flowfield and surface properties can be obtained. Secondly, the Maxwell type surface assumes $\alpha^*$ percent of reflections are diffuse and $1-\alpha^*$ percent reflections are specular. For example, in most engineering applications, $\alpha^* >80\%$ is a good approximation. The related macroscopic properties are different from those for completely diffusely or specularly reflective surfaces. However, the derivation procedures are based on the gaskinetic theory as well, in the same vein as those for the diffusely and specularly reflective surfaces presented in this paper.


}

\section{CONCLUSIONS}
\label{sec:conclusions}
This paper addressed the fundamental problem of free molecular flow through a diffusely or specularly reflective planar micro-channel. Theoretical and numerical investigations are performed and compared. The gaskinetic theory is used to derive the theoretical results, and DSMC simulations are performed to validate those results. Essentially identical results by the two methods indicate that the work is solid. 

For surfaces with diffuse reflections, virtual number densities at the plate surfaces are introduced, and then different VDFs and velocity phases for points at the plate surfaces and inside flowfield are properly constructed. By integrating the VDFs in the velocity phases according to the gaskinetic theory, surface and flowfield properties are obtained. The mass flow rate can be approximated by using the mass flux at the channel center point, and the result includes four non-dimensional parameters: the aspect ratio, density ratio, and two temperature ratios. Different from many past investigations, this work illustrates that the surface temperatures do need to be considered in the mass flow rate for collisionless gaseous micro-channel flows. 

For channels with specularly reflective surfaces, the VDFs at the plate surface and inside the flowfield are identical and the final expressions are explicit and simple, only involving two non-dimensional parameters. The plate surface temperatures and the aspect ratio do not have any influence. No matter how long or high the channel is, and how hot the plate surfaces are, there is no difference in the final properties. 




\section*{DATA AVAILABILITY STATEMENT}
The data that support the findings of this study are available from the corresponding author upon reasonable request.

\section*{APPENDIX}
Several useful integrals:
\begin{equation*}
\begin{array}{rll}
      &\int_{\pm a}^{\infty } e^{- r^2}  dr = \frac{\sqrt{\pi}}{2}  \left[ 1 \mp \mbox{erf}(  a)  \right]; & 
      \int_{\pm a}^{\infty }  r e^{- r^2}  dr =  \frac{1}{2} e^{- a^2}; \\
      &\int_{\pm a}^{\infty }  r^2 e^{- r^2} dr  =     \frac{\sqrt{\pi} }  {4} \left[ 1 \mp \mbox{erf}(  a)  \right] \pm \frac{a}{2} e^{- a^2}; 
      &\int_{\pm a}^{\infty }  r^3 e^{- r^2} dr  =  \frac{1+  a^2}{2} e^{- a^2};   \\
  &     \int_{\pm a}^{\infty }  r^4 e^{- r^2} dr  =  \frac{3\sqrt{\pi}}{8}   \left[ 1 \mp \mbox{erf}( a)  \right]  \pm   \left( \frac{3a}{4}+ \frac{a^3}{2} \right) e^{- a^2};  &
      \int_{\pm a}^{\infty }  r^5 e^{- r^2} dr  =  \left( 1+ a^2 + \frac{a^4}{2}   \right) e^{- a^2}.
\end{array}
\end{equation*}

\reftitle{REFERENCES}


\begin{thebibliography}{999}

\bibitem{Knudsen}
M. Knudsen, ``The law of molecular flow and viscosity of gases moving through tubes,'' Ann. Phys. {\bf 28}, 75  (1909). 

\bibitem{Gaede}
W. Gaede, ``Die äußere reibung der gase,'' Ann. Phys. {\bf 41}, 289 (1913). 

\bibitem{Smoluchowski}
M. V. Smoluchowski, Ann. Phys. {\bf 33}, 1559 (1910). 

\bibitem{Present}
W. C. Pollard,  and R. D. Present, ``On gaseous self-diffusion in long capillary tubes,'' Phys. Rev. {\bf 73}, 762 (1948). 

\bibitem{Takao}
K. Takao,  Trans. Jap. Soc. Aeronaut. Space Sci. {\bf 4}, 82 (1961).

\bibitem{Cercignani1} 
C. Cercignani, and A. Daneri, ``Flow of a rarefied gas between two parallel plates,''  J. Appl. Phys. {\bf 14}, 3509 (1963). 

\bibitem{Cercignani2}
C. Cercignani,  ``Rarefied gas flow through long slots,''  J. Appl. Math. Phys. (ZAMP) {\bf 30} 943-951 (1979).

\bibitem{DeMarcus1}
W. C. DeMarcus, ``The problem of Knudsen flow. Part 3: solutions for one-dimensional systems,''  Oak Ridge Gaseous Diffusion Plant Report, Union Carbide Nuclear Company (Oak Ridge, Tennessee, 1957). 

\bibitem{DeMarcus2}
W. C. DeMarcus, ``The problem of Knudsen flow. Part IV: specular reflection,''  Oak Ridge Gaseous Diffusion Plant Report, Union Carbide Nuclear Company (Oak Ridge, Tennessee, 1957). 

\bibitem{DeMarcus3}
W. C. DeMarcus, ``The problem of Knudsen flow. Part VI: tortuosity,'' Oak Ridge Gaseous Diffusion Plant Report, Union Carbide Nuclear Company (Oak Ridge, Tennessee, 1957). 

\bibitem{Clausing1} 
P. Calusing, Ann. Physik {\bf 12} (5), 961 (1932). Also thesis, Leiden 1928. 

\bibitem{Clausing2}
P. Clausing, ``The flow of highly rarefied gases through tubes of arbitrary length,'' J. Vac. Sci. Technol. A. {\bf 8}, 636 (1971).

\bibitem{Berman}
A. S. Berman, ``Free molecule transmission probabilities,'' J. Appl. Phys.  {\bf 36}, 3356 (1964). 

\bibitem{Bermanerror}
A. S. Berman, ``Erratum: free molecule transmission probabilities,'' J. Appl. Phys.  {\bf 37}, 4598 (1966). 

\bibitem{Dayton}
B. B. Dayton, ``Gas flow in vacuum systems,'' J. Vac. Sci. Technol. A {\bf 9}, 243 (1972). 

\bibitem{Raghuraman}
P. Raghuraman, {\em Kinetic Theory Analysis of Rarefied Gas Flow Through Finite Length Slots}, NASA-CR-124069, Report No. FM-72-1, University of California, Berkly (1972).

\bibitem{Santeler}
D. J. Santeler, ``New concepts in molecular gas flows,'' J. Vac. Sci. Technol. A. {\bf 4}, 338 (1986).

\bibitem{Steckelmacher}
W. Steckelmacher, ``Knudsen flow 75 years on: the current state of the art for flow of rarefied gases in tubes and systems,'' Rep. Prog. Phys. {\bf 49} (10) (1986). 

\bibitem{bird}
G. A. Bird, {\em Molecular Gas Dynamics and Direct Simulation of Gas Flows} (Clarendon Press, Oxford, 1994).

\bibitem{Moran}
M. Wang, and Z. Li, ``Simulations for gas flows in micro-geometries using the direct simulation Monte Carlo method,'' Int. J. Heat Fluid {\bf 25} pp. 975-985 (2004).

\bibitem{Roohi2}
E. Roohi, M. Darbandi, and V. Mirjalili, ``DSMC solution of subsonic flow through micro-nano scale channels,'' J. Heat Trans. {\bf 131} (9), 092402 (2009).

\bibitem{IP}
C. Shen, ``Statistical simulation of low-speed rarefied gas flows,'' J. Comput. Phys. {\bf 167} (2), 393-412 (2001). 

\bibitem{IPCai}
C. Cai, I.D. Boyd, and J. Fan, ``Direct simulation method for low-speed micro-channel flows,'' J. Thermophys.  Heat Trans. {\bf 14} (2), 368–378 (2002).

\bibitem{roohi}
M. Darbandi, and  E. Roohi, ``DSMC simulation of subsonic flow through nano channels and micro/nano backward-facing steps,'' Int. Commun. Heat. Mass. {\bf 38}, 1443–1448 (2011).

\bibitem{Reese}
J. M. Reese, M. A. Gallis, and D. A. Lockerby, ``New directions in fluid dynamics: non-equilibrium aerodynamic and micro-system flows,'' Philos. Trans. R. Soc. Lond. (2003).

\bibitem{Cercignani2004}
C. Cercignani, M. Lampis, and S. Lorenzani, ``Variational approach to gas flows in micro-channels,''  Phys. Fluids {\bf 16}, 3426 (2004).

\bibitem{Titareva}
V. A. Titarava, and E. M. Shakhov, ``Efficient method for computing rarefied gas flow in a long finite plane channel,''  Comput. Math. Math. Phys. {\bf 52} (2), 269–284. (2012)

\bibitem{Sharipov}
S. Pantazis, D. Valougeorgis, and F. Sharipov, ``End corrections for rarefied gas flows through capillaries of finite length,'' Vacuum {\bf 96}, 26-29 (2013). 


\bibitem{Takata1} 
S. Takata, H. Sugimoto, and S. Kosuge,  ``Gas separation by means of the Knudsen compressor,'' Eur. J. Mech. B/Fluids {\bf 26}, 155–181 (2007).

\bibitem{Takata2}
S. Kosuge, and S. Takata, ``Database for flows of binary gas mixtures
through a plane microchannel,''  Eur. J. Mech. B/Fluids {\bf 27}, 444-465 (2008). 


\bibitem{ye}
J. Ye, J. Yang, J. Zheng, P. Xu, and C. Lam, ``Effects of wall temperature on the heat and mass transfer in micro-channels using the DSMC method,'' 4th IEEE-NEMS, Shenzhen, pp. 666-671 (2009).

\bibitem{cshen}
C. Shen, ``Gas flow in micro channels - experimental, computational and kinetic-theoretical investigations,''  Micro. Nano. Systems {\bf 1} (3), pp. 226-233 (2009).  Bentham Science Publishers.

\bibitem{Livesey}
R. G. Livesey,  ``Method for calculation of gas flow in the whole pressure regime through ducts of any length,''  J. Vacu. Sci. Technol. A {\bf 19} (4), 1674-1678 (2001).








\bibitem{Reese2}
N. Dongari, Y.H. Zhang, and J.M. Reese, ``Modeling of Knudsen layer effects in micro/nanoscale gas flows,'' J. Fluids Engineering-Trans {\bf 133} (7) (2011).

\bibitem{Sliu}
S. Liu, C.W. Zhong, and J. Bai,  ``Unified gas-kinetic scheme for microchannel and nanochannel flows,'' Comput. Math. Appl. {\bf 69} (1), 41-57 (2015).

\bibitem{Duan}
Y.D. Yuan,  and S. Rahman, ``Extended application of Lattice Boltzmann Method to rarefied gas flow in micro-channels,'' Physicia A {\bf 463}, 25-36 (2016).

\bibitem{zhihui_Li}
S.M Hou,  Z.H. Li,  X.Y. Jiang, and S. Zeng, ``Numerical study on two-dimensional micro-channel flows using the gas-kinetic unified algorithm,'' Commu. Comput. Phys. {\bf 23}, 1393-1414 (2018).

\bibitem{Srinivasan}
K. Srinivasan,  ``A comprehensive experimental and numerical study on gas flow through micro-channels from slip to free-molecular regimes,'' J. Micromech. Microeng. {\bf 28}, 095006 (2018).


\bibitem{enclosure}
C. Cai,  and D. D. Liu, ``Collisionless gas flows I: inside arbitrary enclosures,'' Phys. Fluids {\bf 20} (6), 067105, (2008).

\bibitem{cai_jsr}
C. Cai, K. Khasawneh, H. Liu, and W. Wei, ``Collisionless gas flows over a cylindrical or spherical object,'' J. Spacecraft  Rockets {\bf 46} (6), 1124-1131 (2009).

\bibitem{aiaaj_cai}
C. Cai, X. He, and K. Zhang, ``Comprehensive studies on rarefied jet and jet impingement flows with gaskinetic methods,''  Commu. Compu. Phys. {\bf 23} (3), 712-741 (2017).

\bibitem{Kogan}
M. N. Kogan, {\em Rarefied Gas Dynamics} (Plenum Press, New York, 1969).

\bibitem{transpiration}
H. Yamaguchi, M. Rojas-Cardenas, P. Perrier, I. Graur, and T. Niimi, ``Thermal transpiration flow through a single rectangular channel,'' J. Fluid Mech. {\bf 744}, 169-182 (2014). 

\bibitem[Liu1(2012)]{GRASP}
H. Liu,  C. Cai, and  C. Zou, ''An object-oriented serial implementation of a DSMC simulation package,''  Comput. Fluids.  {\bf 57}, 66-75 (2012).

\end{thebibliography}
\end{document}